\documentstyle[12pt,epsf]{article}

\catcode`\@=11
\long\def\@makefntext#1{
\protect\noindent \hbox to 3.2pt {\hskip-.9pt  
$^{{\ninerm\@thefnmark}}$\hfil}#1\hfill}		%CAN BE USED 

\def\@makefnmark{\hbox to 0pt{$^{\@thefnmark}$\hss}}  %ORIGINAL 
	
\def\ps@myheadings{\let\@mkboth\@gobbletwo
\def\@oddhead{\hbox{}
\rightmark\hfil\ninerm\thepage}   
\def\@oddfoot{}\def\@evenhead{\ninerm\thepage\hfil
\leftmark\hbox{}}\def\@evenfoot{}
\def\sectionmark##1{}\def\subsectionmark##1{}}

\renewcommand{\thefootnote}{\fnsymbol{footnote}}

\newcounter{sectionc}\newcounter{subsectionc}\newcounter{subsubsectionc}
\renewcommand{\section}[1] {\vspace*{0.6cm}\addtocounter{sectionc}{1} 
\setcounter{subsectionc}{0}\setcounter{subsubsectionc}{0}\noindent 
	{\normalsize\bf\thesectionc. #1}\par\vspace*{0.4cm}}
\renewcommand{\subsection}[1] {\vspace*{0.6cm}\addtocounter{subsectionc}{1} 
	\setcounter{subsubsectionc}{0}\noindent 
	{\normalsize\it\thesectionc.\thesubsectionc. #1}\par\vspace*{0.4cm}}
\renewcommand{\subsubsection}[1]
{\vspace*{0.6cm}\addtocounter{subsubsectionc}{1}
	\noindent {\normalsize\rm\thesectionc.\thesubsectionc.\thesubsubsectionc. 
	#1}\par\vspace*{0.4cm}}

\newcounter{appendixc}
\newcounter{subappendixc}[appendixc]
\newcounter{subsubappendixc}[subappendixc]

\renewcommand{\appendix}[1] {\vspace*{0.6cm}
        \refstepcounter{appendixc}
        \setcounter{figure}{0}
        \setcounter{table}{0}
        \setcounter{equation}{0}
        \renewcommand{\thefigure}{\Alph{appendixc}.\arabic{figure}}
        \renewcommand{\thetable}{\Alph{appendixc}.\arabic{table}}
        \renewcommand{\theappendixc}{\Alph{appendixc}}
        \renewcommand{\theequation}{\Alph{appendixc}.\arabic{equation}}
        \noindent{\bf Appendix \theappendixc #1}\par\vspace*{0.4cm}}

\def\abstracts#1{{
	\centering{\begin{minipage}{12.2truecm}\footnotesize\baselineskip=12pt\noindent
	\centerline{\footnotesize ABSTRACT}\vspace*{0.3cm}
	\parindent=0pt #1
	\end{minipage}}\par}} 

\renewenvironment{thebibliography}[1]
	{\begin{list}{\arabic{enumi}.}
	{\usecounter{enumi}\setlength{\parsep}{0pt}
\setlength{\leftmargin 1.25cm}{\rightmargin 0pt}
	 \setlength{\itemsep}{0pt} \settowidth
	{\labelwidth}{#1.}\sloppy}}{\end{list}}

\newcounter{itemlistc}
\newcounter{romanlistc}
\newcounter{alphlistc}
\newcounter{arabiclistc}

\newcommand{\fcaption}[1]{
        \refstepcounter{figure}
        \setbox\@tempboxa = \hbox{\footnotesize Fig.~\thefigure. #1}
        \ifdim \wd\@tempboxa > 6in
           {\begin{center}
        \parbox{6in}{\footnotesize\baselineskip=12pt Fig.~\thefigure. #1}
            \end{center}}
        \else
             {\begin{center}
             {\footnotesize Fig.~\thefigure. #1}
              \end{center}}
        \fi}

\newcommand{\tcaption}[1]{
        \refstepcounter{table}
        \setbox\@tempboxa = \hbox{\footnotesize Table~\thetable. #1}
        \ifdim \wd\@tempboxa > 6in
           {\begin{center}
        \parbox{6in}{\footnotesize\baselineskip=12pt Table~\thetable. #1}
            \end{center}}
        \else
             {\begin{center}
             {\footnotesize Table~\thetable. #1}
              \end{center}}
        \fi}

\def\@citex[#1]#2{\if@filesw\immediate\write\@auxout
	{\string\citation{#2}}\fi
\def\@citea{}\@cite{\@for\@citeb:=#2\do
	{\@citea\def\@citea{,}\@ifundefined
	{b@\@citeb}{{\bf ?}\@warning
	{Citation `\@citeb' on page \thepage \space undefined}}
	{\csname b@\@citeb\endcsname}}}{#1}}

\newif\if@cghi
\def\cite{\@cghitrue\@ifnextchar [{\@tempswatrue
	\@citex}{\@tempswafalse\@citex[]}}
\def\citelow{\@cghifalse\@ifnextchar [{\@tempswatrue
	\@citex}{\@tempswafalse\@citex[]}}
\def\@cite#1#2{{$\null^{#1}$\if@tempswa\typeout
	{IJCGA warning: optional citation argument 
	ignored: `#2'} \fi}}

 1
 1
 1

\font\ninerm=cmr9

\begin{document}
\hfill FERMILAB--CONF--96--032
%\hfill 7 February 1996 \\

\vspace{3.0cm}

\centerline{\Large\bf Masses and Lifetimes of B Hadrons}

\vspace{1.0cm}

\centerline{\large I.~JOSEPH KROLL}

\vspace{0.5cm}

\centerline{\it Fermilab}
\centerline{\it P.~O.~Box 500}
\centerline{\it Batavia, IL, 60510, U.S.A.}
\centerline{\it E-mail: kroll@fnal.gov}

\vspace{1.0cm}

\centerline{\large\bf Abstract}

\vspace{1.0cm}

\noindent
The latest measurements of the masses and lifetimes of weakly decaying
$B$~hadrons from experiments at $e^+e^-$ and $p\bar{p}$ colliders are
presented. These measurements include the lifetimes of the
$\bar{B}^0$, $\bar{B}^0_s$, $B^-$, and $b$~baryons, as well as 
searches for the $B_c$~meson. The observation of $B^*$, p-wave B~mesons
($B^{**}$), and excited $b$~baryons using inclusive and exclusive $B$~hadron
reconstruction are discussed. Results on $b$~quark flavour tagging are given.

\vspace{3.0cm}

\centerline{\it to appear in the proceedings of the}
\centerline{\it XVII International Symposium on Lepton--Photon Interactions}
\centerline{\it Beijing, 10--15 August 1995}
%\vfill
\pagebreak
\pagestyle{plain}
\pagenumbering{arabic}
\centerline{\normalsize\bf MASSES AND LIFETIMES OF B HADRONS}
%\vfill
%\vspace*{0.6cm}
\centerline{\footnotesize I.~JOSEPH KROLL}
\baselineskip=13pt
\centerline{\footnotesize\it Fermilab}
\centerline{\footnotesize\it P.~O.~Box 500}
\baselineskip=12pt
\centerline{\footnotesize\it Batavia, IL, 60510, U.S.A.}
\centerline{\footnotesize E-mail: kroll@fnal.gov}

%\vfill
\vspace*{0.9cm}
\abstracts{
The latest measurements of the masses and lifetimes of weakly decaying
$B$~hadrons from experiments at $e^+e^-$ and $p\bar{p}$ colliders are
presented. These measurements include the lifetimes of the
$\bar{B}^0$, $\bar{B}^0_s$, $B^-$, and $b$~baryons, as well as 
searches for the $B_c$~meson. The observation of $B^*$, p-wave B~mesons
($B^{**}$), and excited $b$~baryons using inclusive and exclusive $B$~hadron
reconstruction are discussed. Results on $b$~quark flavour tagging are given.}
%%%\input lp_main.tex
%\vspace*{0.6cm}
\normalsize\baselineskip=15pt
\setcounter{footnote}{0}
\renewcommand{\thefootnote}{\alph{footnote}}
\section{Introduction}

The masses and lifetimes of $B$~hadrons are fundamental properties of these
particles and are needed to determine other quantities such as
$V_{cb}$\cite{ref:tomasz}, as well as $\Delta m_d$ and
$\Delta m_s$\cite{ref:saulan}.
%The masses and lifetimes of $B$~hadrons are fundamental properties of these
%particles and are, therefore, basic experimental quantities.
%They are needed, furthermore, to determine other experimental parameters
%such as $V_{cb}$\cite{ref:tomasz} and $\Delta m_d$ and
%$\Delta m_s$\cite{ref:saulan}.
Their values can be used to test theoretical models of heavy quark
decay\cite{ref:matthias}. For example, the $1/m_Q$ expansion in QCD
predicts lifetime differences for the different species of
$B$~hadrons\cite{ref:bigi}. Heavy Quark Effective Theory (HQET) predicts
that the mass differences between excited and ground states
({\it e.g.}~between $B$ and $B^*$) should be independent of the
heavy quark mass, $m_Q$, as $\Lambda_{QCD}/m_Q\rightarrow 0$.
If these models do not predict the experimental observables,
then perhaps the models are not as reliable as we believe.
Finally, the large top~quark mass implies that the top~quark decays
weakly before it hadronizes. The $b$~quark is the heaviest quark with which
we can study these heavy flavour models.

The experimental results presented in this review come from $e^+e^-$ collisions
at the $Z^0$ resonance (LEP at CERN and SLC at SLAC) and from
$p\bar{p}$ collisions at $\sqrt{s}=1.8$~TeV (the Tevatron at Fermilab).

On the $Z^0$ resonance, $b\bar{b}$ production is 22\% of the hadronic
cross~section\cite{ref:renton}.
The experimental trigger efficiency and selection efficiency
for hadronic events are very close to 100\%. The $B$~hadrons have a large
Lorentz boost ($\beta\gamma$): the average value is 6 ({\it i.e.}~a momentum
of 30 GeV/$c$), and the mean distance travelled by a $B$~hadron is 3~mm. 

At the Tevatron, the inclusive $B$~hadron cross~section is
$37\mu$b\cite{ref:ppbarbcross}, which is very large, but
the inelastic cross~section\cite{ref:inelast} is three orders of magnitude
larger. Specialized triggers are required to extract the $B$~hadron signal.
The triggers used in the results presented in this review are 
(1) a dimuon trigger that triggers on the process
$B\rightarrow J/\psi+X, J/\psi\rightarrow\mu^+\mu^-$, and
(2) an inclusive lepton trigger, which is based on the semileptonic decays
of $B$~hadrons:
$B\rightarrow\ell\nu X$, $\ell=e$ or $\mu$. The average transverse momentum
(with respect to the colliding beam axis)
of the $B$~hadrons from the dimuon trigger is 10 GeV/$c$, and the average
transverse momentum from the inclusive lepton trigger is 20 GeV/$c$.
Despite the much larger center-of-mass energy, the $B$~hadrons studied at
the Tevatron are softer than the $B$~hadrons produced on the $Z^0$~resonance.

The data samples used for the results presented in this review are summarized
in Table~\ref{tab:data_samples}. This Table also summarizes the characteristics
of the detectors and accelerators crucial to the lifetime measurements,
namely the silicon strip and pixel vertex detectors and the beam profiles.
The small beam size and close proximity of the first layer of the vertex
detector allow SLD to make competitive lifetime measurements despite
a much smaller data sample than the LEP experiments.

\begin{table}[h] \centering
\tcaption{A summary of the data samples used for the results presented
in this review. The Table also includes information about the silicon
strip and pixel vertex detectors and the sizes of the beam envelopes
at the different accelerators. The data sample size for LEP is the range
of sample sizes of the four experiments. The SLD pixel detector has four
layers, but the average number of hits per track is 2.3. The coordinates
are cylindrical, with the $z$~axis corresponding to the beam line.}
\label{tab:data_samples}
\small
\begin{tabular}{||c|c|c|c||}\hline\hline
Detector & LEP & SLD & CDF \\
\hline
Data Sample & $3.1-3.7\times10^6$ & $1.5\times10^5$ & 120 pb$^{-1}$ \\
            & $Z^{0}\rightarrow$ hadrons & $Z^{0}\rightarrow$ hadrons &     \\
\hline
Vertex      & Silicon strip    & CCD (Pixel)      & Silicon Strip \\
Detector    & $r\phi$ and $rz$ & $r\phi$ and $rz$ & $r\phi$ only \\
\hline
Beam Size   &                 &            &               \\
$\sigma_x$  & 100--160 $\mu$m & 2 $\mu$m   & 35--40 $\mu$m \\
$\sigma_y$  & 5--10 $\mu$m    & 1 $\mu$m   & 35--40 $\mu$m \\
$\sigma_z$  & 7 mm            & 0.7 mm     & 30 cm         \\
\hline
vertex det. &      &        &      \\
inner       & 6 cm & 2.5 cm & 3 cm \\
radius      &      &        &      \\
\hline
\# layers   & 2--3 & 4      & 4    \\
\hline\hline
\end{tabular} 
\end{table}

\section{Masses and Lifetimes of Weakly-decaying $B$~Hadrons}

There are four weakly decaying mesons containing $b$~quarks:
$\bar{B}^0$ ($b\bar{d}$), $B^-$ ($b\bar{u}$), $\bar{B}^0_s$ ($b\bar{s}$),
and $B^-_c$ ($b\bar{c}$); and four predicted weakly decaying baryons:
$\Lambda^0_b$ ($bud$), $\Xi^0_b$ ($bsu$), $\Xi^-_b$ ($bsd$),
and $\Omega^-_b$ ($bss$).\footnote{All references to particles
containing a $b$~quark also imply the equivalent particles containing a
$\bar{b}$~quark, and all decay sequences imply the charge conjugate process
as well.}\ \
Only the $\bar{B}^0$, $B^-$, $\bar{B}^0_s$,
and $\Lambda^0_b$ have been firmly established experimentally.
In the spectator model of the decay of a hadron containing a $b$~quark,
the lifetime is given by
$$\Gamma=1/\tau=\frac{G_F^2 m_b^5}{192\pi^3}
\cdot|V_{bc}|^2\times {\cal F}_{ps},$$
where ${\cal F}_{ps}$ is a phase-space factor
(there is also a term with $|V_{bu}|^2$, which is small and has been neglected).
All $B$~hadrons are predicted to have equal lifetimes.
These hadrons also have equal semileptonic branching fractions, since the
partial width $\Gamma_{\ell}=\Gamma(b\rightarrow\ell\nu c)$ is equal for all
$B$~hadrons, and therefore $B(b\rightarrow\ell\nu c)$ = $\Gamma_{\ell}/\Gamma$
is the same for all $B$~hadrons.

The spectator model failed in the prediction of the equality of the
lifetimes of the charmed hadrons:
$$\tau(D^+)\approx2.5\tau(D^0)\approx2.5\tau(D_s)\approx5.0\tau(\Lambda_c).$$
These lifetime differences are attributed to nonspectator effects such as
final state interference, annihilation diagrams, and helicity suppression.
Applying these same ideas to $B$~hadrons results in the lifetime hierarchy:
$$\tau(\Lambda_b)<\tau(\bar{B}^0)\approx\tau(\bar{B}^0_s)<\tau(B^-).$$
A QCD expansion in powers of $1/m_b$ yields quantitative
predictions\cite{ref:bigi}:
$$\frac{\tau(B^-)}{\tau(\bar{B}^0)}=1+0.05\cdot [\frac{f_b}{200 MeV}]^2,$$
$$\frac{\tau(\bar{B}^0_s)}{\tau(\bar{B}^0)}\approx 1,$$ 
$$\frac{\tau(\Lambda_b)}{\tau(\bar{B}^0)}\approx 0.9,$$
where $f_b$ is the $B$~meson form factor (expected to be 200 to 250 MeV). 

There have been two types of $B$~hadron lifetime measurements:
(1) average and (2) species specific. The average is the lifetime of the
produced mixture of weakly decaying $B$~hadrons. The usual assumption about this
mixture is 40\% $B^-$, 40\% $\bar{B}^0$, 12\% $\bar{B}^0_s$ and 8\% $b$~baryons.
The species specific lifetimes require a fairly pure signature of
a specific $B$~hadron, {\it e.g.}~$\bar{B}^0\rightarrow J/\psi \bar{K}^{*0}$.

Two methods have been employed to make these lifetime measurements. The first is
based on the signed impact parameter $\delta$, which is illustrated in
Figure~\ref{fig:delta}. The Figure illustrates a $B$~hadron, which is produced
at the collision point, propagates away from the point where it is produced,
and then decays. The reconstructed trajectories of charged particles coming
from the $B$~hadron decay are extrapolated back to the collision point.
The distance of closest approach of these extrapolated trajectories to this
collision point is the impact parameter. 
It can be reconstructed in two-dimensions (the plane perpendicular to the
beam line) or three-dimensions.
%It is a one-dimensional or
%two-dimensional quantity, depending on whether the analysis is performed
%in two-dimensions (the plane perpendicular to the beam line) or
%three-dimensions.
%It can be a two-dimensional or three-dimensional quantity.
The impact parameter has a positive (negative) sign if the extrapolated
track trajectory crosses the $B$~hadron flight direction
in front of (behind) the collision point. In the example depicted in
the Figure, the sign is positive. A negative impact parameter can arise
due to either a decay product that goes backwards or due to resolution
effects, either in the determination of the trajectory of the decay product
or in the determination of the $B$~hadron flight direction. The average impact
parameter is proportional to the lifetime of the $B$~hadron. The advantage
of using the impact parameter is that it is fairly insensitive to the
$B$~hadron boost: a $B$~hadron with a larger Lorentz boost will travel farther,
but the decay products come out at a smaller angle, leaving $\delta$ unchanged.
To extract the lifetime, a Monte Carlo model is used to reproduce the observed
impact parameter distribution as a function of the $B$~hadron lifetime.
The sign of the impact parameter is important for the Monte Carlo modelling:
since the negative part of the impact parameter distribution is dominated
by resolution effects, it provides a means to control the resolution in
the Monte Carlo model. Typically impact parameter measurements are made using
leptons (electrons or muons) from $B$~hadron semileptonic decays; in fact,
the signed impact parameter of leptons was the method employed
to make the first measurements\cite{ref:pep} of the $b$ lifetime.

\begin{figure}
\begin{picture}(100,300)(-100,-50)
\put (0,0){\circle*{5}}
%\put (0,0){\line(0,1){100}}
%\put (0,0){\line(1,0){100}}
%\put (0,0){\vector(1,1){100}}
\multiput(0,0)(2,2){51}{\circle*{1}}
\put (100,100){\circle*{5}}
\thicklines
\put (100,100){\vector(4,3){150}}
\thinlines
\put (100,100){\vector(1,0){100}}
\put (100,100){\vector(3,1){94.9}}
\put (100,100){\vector(2,3){55.5}}
\put (100,100){\vector(1,2){44.7}}
%\put (100,100){\line(-4,-3){112}}
\multiput(-12,16)(22.4,16.8){7}{\line(4,3){11.2}}
%\multiput(100,100)(-3.2,-2.4){35}{\circle*{1}}
\put (0,0){\vector(-3,4){12}}
\put (-12,16){\vector(3,-4){12}}
\put (-15,0){$\delta$}
\put (35,0){\vector(-1,0){28}}
\put (37,0){$B$ production point}
\put (65,100){\vector(1,0){28}}
\put (-10,100){$B$ decay point}
\end{picture}
\fcaption{The definition of the signed-impact parameter $\delta$.}
\label{fig:delta}
\end{figure}

The second method for measuring lifetimes is based on the decay length,
%which is illustrated in Figure~\ref{fig:decayl}.
which is the distance from the $B$~hadron production point to the $B$~hadron
decay point. The decay length $L$ is related to the lifetime $c\tau$ by the
Lorentz boost $\beta\gamma$:

$$L=\beta\gamma c\tau.$$

Unlike the impact parameter method, it is necessary to know the boost value.
In the case of a fully reconstructed $B$~hadron decay, the determination
of the boost value is straightforward:
$$\beta\gamma=p_B/m_B,$$
where $p_B$ is the $B$~hadron momentum and $m_B$ is the $B$~hadron mass.

%The utilization of precise silicon tracking detectors made the reconstruction
%of the decay length possible. 

\subsection{Average Lifetime of $B$~Hadrons}

Measurements of the average $B$~hadron lifetime have been made using both
the impact parameter method and the decay length method. The results are
summarized in Table~\ref{tab:avg}.

\begin{table}[h] \centering
\tcaption{Summary of average $B$~hadron lifetimes. The first error is the
statistical error, and the second error is the systematic error.
For impact parameter methods, 2D and 3D indicate whether the analysis
is performed in two-dimensions or three-dimensions.}
\label{tab:avg}
\small
\begin{tabular}{||c|c|c|c|c||}\hline\hline
Experiment & Method & Data Set & Result (psec) & Reference \\
\hline
ALEPH  & lepton $\delta$ (3D) & 91--93 & $1.533\pm0.013\pm0.022$ & 
\cite{ref:aleph_avgb_lepton} \\
\hline
ALEPH  & dipole               & 91     & $1.511\pm0.022\pm0.078$ &
\cite{ref:aleph_avgb_dipole} \\
\hline
DELPHI & hadron $\delta$ (2D) & 91--92 & $1.542\pm0.021\pm0.045$ &
\cite{ref:delphi_avgb_had}   \\
\hline
DELPHI & hadron vertex        & 91--93 & $1.600\pm0.010\pm0.028$ &
\cite{ref:delphi_eps0751}    \\
\hline
L3     & lepton $\delta$ (2D) & 90--91 & $1.535\pm0.035\pm0.028$ &
\cite{ref:l3_avgb_lepton}    \\
\hline
OPAL   & lepton $\delta$ (2D) & 90--91 & $1.523\pm0.034\pm0.038$ &
\cite{ref:opal_avgb_lepton}  \\
\hline
SLD    & hadron vertex        & 93     & $1.564\pm0.030\pm0.036$ &
\cite{ref:sld_avgb}          \\
\hline
CDF    & $J/\psi$ vertex      & Run 1a & $1.46 \pm0.06 \pm0.06 $ &
\cite{ref:cdf_avgb_jpsi}     \\
\hline\hline
\end{tabular} 
\end{table}

Four measurements based on impact parameter ($\delta$) are reported.
Three are based on the impact parameter of leptons.
The lepton selections yield samples that
have a $B$~hadron purity between 85\% and 95\%. The dominant systematic errors
are due to the modelling of $b$~quark fragmentation and $B$~hadron decay,
the understanding of the impact parameter resolution, and the background
shape. 

Four measurements based on reconstructed decay vertices are reported.
DELPHI and SLD reconstruct the decay length of vertices found in
hadronic $Z^0$ decays. ALEPH uses the so-called dipole method, which measures
the distance between the $B$ vertex and the $\bar{B}$ vertex.
This method does not rely on the determination of the production
point of the $B$ and $\bar{B}$. Finally, CDF uses the vertex reconstructed
from $J/\psi\rightarrow \mu^+\mu^-$ coming from $B\rightarrow J/\psi+X$
decays (ALEPH, DELPHI, and OPAL also have results coming from this signature,
but the statistics are much lower than the CDF result).

These average lifetime measurements must be compared with caution. The different
methods may select different mixtures of $B$~hadrons and therefore might not
be measuring the same average. For example, the measurements based on
lepton impact parameter are an average of the mixture of $B$~hadrons produced
times their semileptonic branching fractions:
$$<\tau>=\sum_{i=species} f(b\rightarrow B_i)B(B_i\rightarrow\ell)\tau(B_i).$$
Average lifetimes based on vertices may have biases favouring neutral or charged
$B$~hadrons depending whether the minimum number of tracks required for a vertex
is even or odd. Because of the potential differences between methods, an
average of the measurements is not given. The two most precise measurements
are from ALEPH\cite{ref:aleph_avgb_lepton} (using lepton impact parameter)
and DELPHI\cite{ref:delphi_eps0751} (using hadronic decay vertices).
The DELPHI systematic error is dominated by uncertainties
in $b$~quark fragmentation and Monte Carlo statistics.
Both measurements have total errors of less than 2\%, but they differ
by 1.7$\sigma$.

\subsection{$\bar{B}^0$ and $B^-$}

%Species specific lifetime measurements have been made using three techniques:

Three types of signatures have been used to measure species specific lifetimes:
\begin{enumerate}
\item{
{\it Fully reconstructed decays:} all decay products of the $B$~hadron are
detected and the $B$~hadron is fully reconstructed. The disadvantage of this
approach is the small branching fractions to particular final states yield
low statistics samples. The advantages are the straightforward determination
of the decay length, the Lorentz boost, and the background
(from mass distribution sidebands).}
\item{
{\it Partially reconstructed decays:} these final states typically consist of
a lepton and a fully reconstructed charmed hadron. The advantage of this
method is significantly larger statistics than fully reconstructed decays,
but the disadvantages include systematics in determining the Lorentz boost,
larger backgrounds, contamination from other species of $B$~hadrons,
and reconstructing the decay length.}
\item{
{\it Inclusive vertex with charge determination:} this technique has 
been used on the $Z^0$~resonance to determine the $\bar{B}^0$ and $B^-$
lifetimes. The $B$~hadron decay vertex is reconstructed, and the charge of the
$B$~hadron is determined from the sum of the charges associated to
the $B$~vertex. Monte Carlo simulation
is needed to determine how well the charged and neutral states are
separated and the effect of $\bar{B}^0_s$ and $b$~baryons.}
\end{enumerate}

Two experiments have reported lifetimes of $\bar{B}^0$ and $B^-$ based on
exclusive decays. CDF exploits a $J/\psi\rightarrow\mu^+\mu^-$ trigger
to reconstruct large samples of $B\rightarrow J/\psi K$. From a data
sample of 67 pb$^{-1}$, they have a signal of $524\pm29$ $B^-$, predominantly
from $B^-\rightarrow J/\psi K^-$, and $285\pm21$ $\bar{B}^0$, predominantly
from $B^0\rightarrow J/\psi K^{*0}$ and $B^0\rightarrow J/\psi K^0_s$.
The ALEPH experiment takes advantage of good particle identification
($dE/dx$) and excellent charged particle momentum resolution to reconstruct
94 $B^-$ candidates and 121 $\bar{B}^0$ candidates from a large number
of final states\cite{ref:aleph_eps0412} using $3\times10^6$ hadronic
$Z^0$ decays from 1991 to 1994. The measured lifetimes are reported
in Table~\ref{tab:b0d} ($\bar{B}^0$) and Table~\ref{tab:bplus} ($B^-$).

Four experiments (ALEPH, DELPHI, OPAL, CDF) have reported lifetimes of
$\bar{B}^0$ and $B^-$ based on partially reconstructed exclusive final states.
The signature is
$\bar{B}^0\rightarrow D^{*+}\ell^-\bar{\nu}$ and 
$\bar{B}^0\rightarrow D^{+}\ell^-\bar{\nu}$ for the neutral $B$~meson, and 
$B^-\rightarrow D^{0} \ell^-\bar{\nu}$ for the charged $B$~meson.

Figure~\ref{fig:dlnu} depicts a specific example of one of these decays:
$\bar{B}^0\rightarrow D^{*+}\ell^-\bar{\nu}$ with
$D^{*+}\rightarrow D^0\pi^+$, $D^0\rightarrow K^-\pi^+$,
and serves to illustrate several important features of lifetime measurements
based on this type of signature.
\begin{itemize}
\item{
The lepton and the charged~K have the same charge. Candidates in which the
lepton and charged~K have opposite charge provide a sample that can be used
to study the combinatorial background.}
\item{
Requirements on kinematic properties of the lepton such as the momentum of the
lepton and the component of the momentum of the lepton perpendicular to the
$B$~hadron momentum ($p^{rel}_t$) are used to suppress backgrounds from 
misidentified leptons and leptons not arising from $B$~hadron semileptonic
decays.}
\item{Since the momentum of the neutrino is not directly measured, the
Lorentz boost of the $B$~hadron must be estimated from the observed
lepton and $D$~meson. This estimation requires input from Monte Carlo
models and the typical resolution on $\beta\gamma$ is
$\sigma_{\beta\gamma}\approx15$\%. Alternatively, on the $Z^0$~resonance,
the energy of the neutrino can be estimated from the observed momentum
imbalance of the event. With this approach, the ALEPH experiment achieves
a resolution on the neutrino energy of $\sigma(E_{\nu})\approx3$ GeV.}
\item{
The $D^0$ is fully reconstructed and is extrapolated back to the point
where it intersects with the lepton to form the $B$ decay vertex.
The $D^0$ decay vertex is measured as well, so the $D^0$ lifetime can
be measured and used to search for biases in the lifetime determination.
The typical resolution on the $D$ decay length is $\sigma(L_D)\sim L_D$.
In contrast, the resolution on the decay length of the $B$~hadron is typically
an order of magnitude less than the decay length itself,
$\sigma(L_B)\sim L_B/10$. Exact modelling of the decay length resolution
is not crucial and is not a dominant systematic error.}
\end{itemize}

The $D^{*+}\ell^-\bar{\nu}$, $D^{+} \ell^-\bar{\nu}$ sample
and the $D^{0} \ell^-\bar{\nu}$ sample do not comprise unique
samples of $\bar{B}^0$ and $B^-$ mesons: there is some cross-contamination
between the two $B$~meson types. There is a small contamination in
the $B^-$ sample from $\bar{B}^0\rightarrow D^{*+}\ell^-\bar{\nu}$ with
$D^{*+}\rightarrow D^0\pi^+$, and the soft pion is not detected.
This contamination is small, because the detection efficiency for the soft
pion is greater than 85\% in all experiments. The more serious source
of cross-contamination comes from semileptonic decays involving $D^{**}$ and
nonresonant $D\pi$. For example, the decay sequence
$\bar{B}^0\rightarrow D^{**+}\ell^-\bar{\nu}$ with
$D^{**+}\rightarrow D^{0}\pi^+$ will be classified as a
``$D^{0}\ell^-\bar{\nu}$'' final state and interpreted as coming from a $B^-$.
%The potential size of this cross-talk can be illustrated using measured
%branching fractions from the $\Upsilon(4S)$
The fraction of semileptonic decays involving a $D^{**}$ and nonresonant $D\pi$
is approximately 30--40\%. In these analyses, the fraction of decays involving
$D^{**}$ is assumed to be $f^{**}=36\pm12$\%. If the ratio of lifetimes
$\tau(B^-)/\tau(B^0)$ is unity, then the
$D^0\ell^-\bar{\nu}$ sample is typically 80--90\% $B^-$ and the
$D^{(*)+}\ell^-\bar{\nu}$ sample is typically 70--80\% $B^0$.
In the future, measurements of $f^{**}$ such as those presented by
the ALEPH\cite{ref:aleph_eps0426} and OPAL\cite{ref:opal_dds} collaborations
will be used to reduce the systematic error due to cross-contamination between
the partially reconstructed $\bar{B}^0$ and $B^-$ samples.

\begin{figure}[h]
\begin{picture}(100,300)(-100,-50)
\put (0,0){\circle*{5}}
\multiput(0,0)(2,2){51}{\circle*{1}}
\put (100,100){\circle*{5}}
\put (100,100){\vector(3,4){100}}
\put (100,100){\vector(1,2){44.7}}
\put (100,100){\vector(1,-2){22.4}}
\multiput(100,100)(3,0){16}{\circle*{1}}
\put (150,100){\circle*{5}}
\put (150,100){\vector(4,3){50}}
\put (150,100){\vector(4,-3){50}}
\put (210,137.5){$K^-$}
\put (210,62.5){$\pi^+$}
\put (210,233.3){$\ell^-$}
\put (144.7,199.4){$\bar{\nu}$}
\put (122.4,45.2){$\pi^+$}
\put (35,0){\vector(-1,0){28}}
\put (37,0){$B$ production point}
\put (65,117.5){\vector(2,-1){28}}
\put (-10,117.5){$B$ decay point}
\put (30,40){\vector(-1,-1){30}}
\put (60,70){\vector(1,1){30}}
\put (35,55){$L_B$}
\put (120,90){$L_D$}
\put (185,100){\vector(-1,0){28}}
\put (187,100){$D^0$ decay point}
\end{picture}
\fcaption{Illustration of the decay$\bar{B}^0\rightarrow D^{*+}\ell^-\bar{\nu}$
with $D^{*+}\rightarrow D^0\pi^+$, $D^0\rightarrow K^-\pi^+$.}
\label{fig:dlnu}
\end{figure}

There are two sources of backgrounds in these partially reconstructed
final states: (1) physics backgrounds from decays such as
$B\rightarrow D^-_sD$, followed by $D^-_s\rightarrow\ell^-\bar{\nu}X$,
and (2) combinatoric background that results from a misidentified lepton
combined with a real $D$ or a real lepton combined with a combinatorial D.
The physics background is suppressed by kinematic requirements on the lepton
and is small. The combinatorial background is suppressed using particle
identification and kinematic requirements, but is often the source of the
largest systematic error.
The lifetime distribution of the background can be determined
from the data by using the $D$ sideband distributions and lepton~$D$
combinations that have the wrong charge correlation.

Table~\ref{tab:dlnusam} reports the signal sizes from the various experiments
and the lifetime measurements are given in Table~\ref{tab:b0d} and
Table~\ref{tab:bplus}. The statistics of the signals from these semi-exclusive
final states are substantially larger than the statistics of the fully
reconstructed decays. 

Table~\ref{tab:b0d} lists one other measurement of $\tau(\bar{B}^0)$
based on partially reconstructed $\bar{B}^0\rightarrow\pi^-_B D^{*+} X$ decays.
This result comes from the ALEPH experiment and is based on the decay chain:
$\bar{B}^0\rightarrow\pi^-_B D^{*+} X$, $D^{*+}\rightarrow D^0\pi^+_D$.
The observed $\pi^-_B\pi^+_D$ system is used to isolate the signal
and to determine the $\bar{B}^0$ momentum and decay point.

\begin{table}[h] \centering
\tcaption{Summary of the lepton plus $D$ signal statistics
(background subtracted)
for the $\bar{B}^0$ and $B^-$ lifetime measurements. The number of $Z^0$'s
means hadronic $Z^0$ decays.}
\label{tab:dlnusam}
\small
\begin{tabular}{||c|c|c|c|c||}\hline\hline
Experiment & Data Sample                  & $\bar{B^0}$ & $B^-$      & Ref. \\
\hline\hline
ALEPH      & $3\times10^6 Z^0$ (91--94)   & $865\pm31$  & $672\pm29$ &
\cite{ref:aleph_eps0412} \\
\hline
DELPHI     & $1.7\times10^6 Z^0$ (91--93) & $309\pm22$  & $377\pm28$ &
\cite{ref:delphi_ld}     \\
\hline
OPAL       & $1.7\times10^6 Z^0$ (91--93) & $697\pm37$  & $292\pm23$ &
\cite{ref:opal_ld}       \\
\hline
CDF        & 20 pb$^{-1}$ Run 1a          & $889\pm35$  & $558\pm35$ &
\cite{ref:cdf_ld}        \\
\hline\hline
\end{tabular} 
\end{table}

DELPHI and SLD report measurements of $\tau(\bar{B}^0)$ and $\tau(B^-)$ 
based on inclusive vertexing with charge determination.
In the DELPHI analysis, $B$ hadrons are selected as jets with a secondary
vertex\footnote{The term ``secondary vertex'' implies that the charged particle
trajectories of the particles in the jet are consistent with originating
from a position (the $B$~hadron decay point) other than the collision point
(the $B$~hadron production point). The vertex may be reconstructed in a plane
or in space. The reconstructed collision point or interaction point is often
referred to as the ``primary vertex.''}\ \
and the sum of the charges of the tracks associated to the
secondary vertex determines the $B$ hadron charge. A minimum of three
tracks must be associated with the secondary vertex, and no attempt
is made to distinguish between the $B$ vertex and the subsequent charmed
particle vertex. This procedure finds 1817 $B$ candidates in a sample of
$1.4\times 10^6$ hadronic $Z^0$ decays. According to Monte Carlo simulation,
these events are $99.1\pm0.3$\% pure in the reaction $Z^0\rightarrow b\bar{b}$,
and 83\% (70\%) of the $B$ hadrons measured as neutral (charged) are
indeed from neutral (charged) $B$~hadrons. The mean lifetime of the
mixture of neutral $B$~hadrons in this data sample is $1.58\pm0.11\pm0.09$~psec
(the first error is statistical and the second error is systematic).
Correcting for the contribution to the sample from $\bar{B}^0_s$
and $\Lambda^0_b$, they find a $\bar{B}^0$ lifetime of
$\tau(\bar{B}^0)=1.63\pm0.14\pm0.13$~psec.
The $B^-$ lifetime is $\tau(B^-)=1.72\pm0.08\pm0.06$~psec.

SLD uses two methods to determine $\tau(\bar{B}^0)$ and $\tau(B^-)$
from a sample of $1.5\times 10^5$ hadronic $Z^0$ decays.
In the first method, they select a sample of high $p$ and $p^{rel}_t$
leptons that come from $B$~hadron semileptonic decay. Next they
try to reconstruct the associated charmed particle from tracks that
are inconsistent with coming from the interaction point.
The method is similar to the method based on the $\ell + D$ signature,
except they do not reconstruct the $D$ meson exclusively. 
The tracks associated with the charmed particle decay are combined
to form a vertex, and the charmed particle is intersected with
the lepton to form the $B$ decay point. The $B$~hadron charge is
determined from the tracks associated to the $B$ decay point.
This method isolates 977 semileptonic $B$ decays, of which
428 are reconstructed as neutral and 549 are reconstructed as charged.
The predicted fractions of $\bar{B}^0$ and $B^-$ in these two samples
are 65.4\% and 70.1\%, respectively. 

The second method employed by SLD is similar to the topological method
of DELPHI, described above. Using track impact parameters they isolate
a subsample of 14\thinspace000 hadronic decays that is 90\% pure in the
process $Z^0\rightarrow b\bar{b}$. Next they associate tracks with
a single secondary vertex in each event hemisphere. As in the DELPHI
analysis, the tracks associated to this vertex come from $B$~hadron decay
and the decay of the associated charmed particle. 
%SLD employs a novel vertexing algorithm that exploits the unique vertexing
%capabilities of the experiment to find vertices in three-dimensions. 
Exploiting the unique vertexing capabilities of SLD, a novel algorithm
is used to reconstruct vertices in three-dimensions;
8685 vertices are reconstructed, with a predicted $B$ purity of exceeding
99\%. These vertices are classified into 3382 neutral $B$ decays and
5303 charged $B$ decays, which are 55.5\% $\bar{B}^0$ and 56.2\% $B^-$,
respectively.
This sample of inclusively reconstructed neutral and charged $B$ decays
is significantly larger than the sample reconstructed with DELPHI
(albeit with poorer separation between charged and neutral $B$~hadrons)
despite starting with an order of magnitude less statistics.
%The overlap between the two SLD methods is large.
The resulting lifetimes are reported in Table~\ref{tab:b0d} ($\bar{B}^0$)
and Table~\ref{tab:bplus} ($B^-$).

\begin{table}[h] \centering
\tcaption{Summary of $\bar{B}^0$ lifetimes. The first error is the
statistical error, and the second error is the systematic error.}
\label{tab:b0d}
\small
\begin{tabular}{||c|c|c|c|c||}\hline\hline
Experiment & Method & Data Set & Result (psec) & Reference \\
\hline
ALEPH  & $\ell + D$           & 91--94 & $1.61\pm0.07\pm0.04$          &
\cite{ref:aleph_eps0412} \\
\hline
ALEPH  & exclusives           & 91--94 & $1.25^{+0.15}_{-0.13}\pm0.05$ &
\cite{ref:aleph_eps0412} \\
\hline
ALEPH  & $\pi^+\pi^-$         & 91--94 & $1.49^{+0.17+0.08}_{-0.15-0.06}$ &
\cite{ref:aleph_eps0412} \\
\hline
DELPHI & $\ell + D$           & 91--93 & $1.61^{+0.14}_{-0.13}\pm0.08$ &
\cite{ref:delphi_ld}     \\
\hline
DELPHI & vertex charge        & 91--93 & $1.63\pm0.14\pm0.13$          &
\cite{ref:delphi_vtx}    \\
\hline
OPAL   & $\ell + D$           & 91--93 & $1.53\pm0.12\pm0.08$          &
\cite{ref:opal_ld}       \\
\hline
\multicolumn{3}{||c|}{LEP Average}      & $1.55\pm0.06$                &
\cite{ref:hgm}           \\
\hline
SLD    & $\ell\ +$ vertex     & 93--95 & $1.60^{+0.15}_{-0.14}\pm0.10$ &
\cite{ref:sld_vtx}       \\
\hline
SLD    & vertex charge        & 93--95 & $1.55\pm0.07\pm0.12$          &
\cite{ref:sld_vtx}       \\
\hline
CDF    & $\ell + D$           & Run 1a & $1.57\pm0.08\pm0.07$          &
\cite{ref:cdf_ld}        \\
\hline
CDF    & excl.~($J/\psi K$)   & Run 1a/1b & $1.64\pm0.11\pm0.06$       &
\cite{ref:cdf_excl}      \\
\hline\hline
\multicolumn{3}{||c|}{World Average}      & 
\multicolumn{2}{|c||}{$1.56\pm0.05$} \\
\hline\hline
\end{tabular} 
\end{table}

\begin{table}[h] \centering
\tcaption{Summary of $B^-$ lifetimes. The first error is the
statistical error, and the second error is the systematic error.}
\label{tab:bplus}
\small
\begin{tabular}{||c|c|c|c|c||}\hline\hline
Experiment & Method & Data Set & Result (psec) & Reference \\
\hline
ALEPH  & $\ell + D$           & 91--94 & $1.58\pm0.09\pm0.04$          &
\cite{ref:aleph_eps0412} \\
\hline
ALEPH  & exclusives           & 91--94 & $1.58^{+0.21}_{-0.18}\pm0.04$ &
\cite{ref:aleph_eps0412} \\
\hline
DELPHI & $\ell + D$           & 91--93 & $1.61\pm0.16\pm0.12$          &
\cite{ref:delphi_ld}     \\
\hline
DELPHI & vertex charge        & 91--93 & $1.72\pm0.08\pm0.06$          &
\cite{ref:delphi_vtx}    \\
\hline
OPAL   & $\ell + D$           & 91--93 & $1.52\pm0.14\pm0.09$          &
\cite{ref:opal_ld}       \\
\hline
\multicolumn{3}{||c|}{LEP Average}      & $1.62\pm0.06$                &
\cite{ref:hgm}           \\
\hline
SLD    & $\ell\ +$ vertex     & 93--95 & $1.49^{+0.11}_{-0.10}\pm0.05$ &
\cite{ref:sld_vtx}       \\
\hline
SLD    & vertex charge        & 93--95 & $1.67\pm0.06\pm0.09$          &
\cite{ref:sld_vtx}       \\
\hline
CDF    & $\ell + D$           & Run 1a & $1.51\pm0.12\pm0.08$          &
\cite{ref:cdf_ld}        \\
\hline
CDF    & excl.~($J/\psi K$)   & Run 1a/1b & $1.68\pm0.09\pm0.06$       &
\cite{ref:cdf_excl}      \\
\hline\hline
\multicolumn{3}{||c|}{World Average}      & 
\multicolumn{2}{|c||}{$1.62\pm0.05$} \\
\hline\hline
\end{tabular} 
\end{table}

The LEP averages reported in Table~\ref{tab:b0d} and Table~\ref{tab:bplus}
have been determined following the methods adopted by the LEP B lifetime
working group\cite{ref:hgm}.
In forming these averages, an attempt is made to standardize
assumptions about aspects of the measurements that are common to a particular
method of measuring a lifetime (for example, the value of $f^{**}$ in the
lepton plus charm measurements).
The world averages in these Tables adopt the same assumptions
used in forming the LEP averages. If instead the individual results are
combined by (1) first adding the statistical and systematic errors of
each individual result in quadrature to get a total error (if the error
is asymmetric, then symmetrize the error by choosing the larger of the
two asymmetric values), and then (2) averaging these individual results
weighing each result by its fractional error ($\sigma_{\tau}/\tau$), the
resulting averages are very similar to the averages reported in the
Tables, and the errors on the averages are only slightly smaller
(by 10\% at most). This agreement is not surprising since the precision of
the individual measurements is dominated by statistical and uncorrelated
systematic errors.

Table~\ref{tab:ratio} summarizes the results on the ratio of lifetimes
$\tau(B^-)/\tau(\bar{B}^0)$.
The world average $\tau(B^-)/\tau(\bar{B}^0)=1.02\pm0.04$ is consistent
with theoretical expectations. 
Many systematic errors that are present in the individual lifetimes
are correlated in this ratio. Due to these correlations, the average
lifetime ratio determined from several experiments must be
based on the average of the ratios determined by each experiment and
not on the ratio of the world averages of $\tau(B^-)$ and $\tau(\bar{B}^0)$.

\begin{table}[h] \centering
\tcaption{Summary of the $B^0/B^+$ lifetime ratio. The first error is the
statistical error, and the second error is the systematic error.}
\label{tab:ratio}
\small
\begin{tabular}{||c|c|c|c|c||}\hline\hline
Experiment & Method & Data Set & Result (psec) & Reference \\
\hline
ALEPH  & $\ell + D$           & 91--94 & $0.98\pm0.08\pm0.02$          &
\cite{ref:aleph_eps0412} \\
\hline
ALEPH  & exclusives           & 91--94 & $1.27^{+0.23}_{-0.19}\pm0.03$ &
\cite{ref:aleph_eps0412} \\
\hline
DELPHI & $\ell + D$           & 91--93 & $1.00^{+0.17}_{-0.15}\pm0.10$ &
\cite{ref:delphi_ld}     \\
\hline
DELPHI & vertex charge        & 91--93 & $1.06^{+0.13}_{-0.11}\pm0.10$ &
\cite{ref:delphi_vtx}    \\
\hline
OPAL   & $\ell + D$           & 91--93 & $0.99\pm0.14^{+0.05}_{-0.04}$ &
\cite{ref:opal_ld}       \\
\hline
\multicolumn{3}{||c|}{LEP Average}      & $1.04\pm0.06$                &
\cite{ref:hgm}           \\
\hline
SLD    & $\ell\ +$ vertex     & 93--95 & $0.94^{+0.14}_{-0.12}\pm0.07$ &
\cite{ref:sld_vtx}       \\
\hline
SLD    & vertex charge        & 93--95 & $1.08^{+0.09}_{-0.08}\pm0.10$ &
\cite{ref:sld_vtx}       \\
\hline
CDF    & $\ell + D$           & Run 1a & $0.96\pm0.10\pm0.05$          &
\cite{ref:cdf_ld}        \\
\hline
CDF    & excl.~($J/\psi K$)   & Run 1a/1b & $1.02\pm0.09\pm0.01$       &
\cite{ref:cdf_excl}      \\
\hline\hline
\multicolumn{3}{||c|}{World Average}      & 
\multicolumn{2}{|c||}{$1.02\pm0.04$} \\
\hline\hline
\end{tabular} 
\end{table}

\subsection{$\bar{B}^0_s$}

There are two new results on $m(\bar{B}^0_s)$ since the last Lepton-Photon
Symposium. OPAL reports\cite{ref:opal_jpsi} a new
$\bar{B}^0_s\rightarrow J/\psi\phi$ candidate,
which combined with their previous candidate yields a mass of
$m(\bar{B}^0_s)=5367\pm15\pm5$ MeV/$c^2$. The first error is the
statistical error, and the second error is the systematic error.
CDF has submitted\cite{ref:cdf_bsmass} the final analysis of the
data collected during Run 1a of the Tevatron Collider (20~pb$^{-1}$).
Based on a signal of $32\pm6$ $\bar{B}^0_s\rightarrow J/\psi\phi$ candidates,
they measure $m(\bar{B}^0_s)=5369.9\pm2.3\pm1.3$ MeV/$c^2$,
which is an improvement of the current value reported in the
Review of Particle Properties\cite{ref:pdg}:
$m(\bar{B}^0_s)=5375\pm6$ MeV/$c^2$.
A similar analysis of $J/\psi K^-$ and $J/\psi K^{*0}$ candidates yields
$m(\bar{B}^0)=5281.3\pm2.2\pm1.4$ MeV/$c^2$,
$m(B^-)=5279.1\pm1.7\pm1.4$ MeV/$c^2$, respectively, yielding a mass
difference of
$\Delta m(\bar{B}^0_s - B)=89.7\pm2.7\pm1.2$ MeV/$c^2$, where ``$B$''
refers to the average of $\bar{B}^0$ and $B^{-}$.

The most common signature used to measure the $\bar{B}^0_s$ lifetime is
opposite-charge $\ell^{-}D^{+}_s$ combinations arising from
$\bar{B}^0_s\rightarrow D^+_s\ell^-\bar{\nu}X$. The most common decay
modes used to reconstruct the $D^+_s$ are $\phi\pi^+$ and $\bar{K}^{*0}K^+$.
The ALEPH experiment uses the additional hadronic modes $\phi\pi\pi\pi$,
$\bar{K}^0_s K^+$, and $\bar{K}^0_s K^{*+}$, as well as the semileptonic
mode $\phi\ell^+\nu$. Table~\ref{tab:dslnusam} summarizes the
$\ell^{-}D^{+}_s$ signals used in the lifetime measurements.
The two physics backgrounds to this $\bar{B}^0_s$ signature are 
\begin{enumerate}
\item{
$B$~meson decays in which the virtual~$W$ emitted by
the $b$~quark forms a $D^-_s$, yielding $B\rightarrow D^-_s D$,
followed by $D\rightarrow\ell^+ \nu X$ yielding a final state with
$D^-_s\ell^+$. This background is suppressed by kinematic requirements
on the lepton, which has a softer $p$ and $p^{rel}_t$ spectrum than
leptons coming from direct $B$ semileptonic decay.}
\item{
Semileptonic decays of $B$~mesons
$B\rightarrow\ell^-\bar{\nu}D^+_s K$,
in which the $c$~quark combines with
the $\bar{s}$~quark of an $s\bar{s}$~quark-pair from the vacuum, and
the spectator antiquark combines with the $s$~quark to form a $K$.
This decay has never been established\cite{ref:argus_cleo}
and is expected to contribute a very
small background. The possible presence of this background 
usually contributes only to the systematic error (the lifetime is not
adjusted).}
\end{enumerate}
There are potential reflections from $B$~meson semileptonic decays that
produce a $D^+$ decay via $D^+\rightarrow \bar{K}^{*0}\pi^+$.
This background can be reduced by using particle identification
to reduce the fraction of pions misidentified as kaons.
Finally, the largest background is combinatoric. This background can be
studied by using the $D^+_s$ sidebands and same-sign $\ell^{\pm}D^{\pm}_s$
combinations.

\begin{table}[h] \centering
\tcaption{Summary of $\ell^{\mp} D^{\pm}_s$ signal statistics
(combinatorial background subtracted) for the $\bar{B}^0_s$
lifetime measurements. The number of $Z^0$'s means hadronic $Z^0$ decays.}
\label{tab:dslnusam}
\small
\begin{tabular}{||c|c|c|c||}\hline\hline
Experiment & Data Sample                  & Signal      &  Ref. \\
\hline\hline
ALEPH      & $3.0\times10^6 Z^0$ (91--94) & $147\pm14$  &
\cite{ref:aleph_bstau}    \\
\hline
DELPHI     & $3.2\times10^6 Z^0$ (91--94) & $ 85\pm13$  &
\cite{ref:delphi_eps0559} \\
\hline
OPAL       & $3.6\times10^6 Z^0$ (91--94) & $ 84\pm13$  &
\cite{ref:opal_bstau}    \\
\hline
CDF        & 20 pb$^{-1}$ Run 1a          & $ 76\pm8$   &
\cite{ref:cdf_bstau}      \\
\hline\hline
\end{tabular} 
\end{table}

Four other signatures have been used to measure the $\bar{B}^0_s$ lifetime.
ALEPH and DELPHI use a $D^+_s$ vertexed with a negatively charged hadron. 
This signature is based on hadronic decays such as
$\bar{B}^0_s\rightarrow D^+_s\pi^-$ or $D^+_s\rho^-$, which presumably have
larger branching fractions than the semileptonic mode used above.
The purity (20--30\%), however, of this signature is much less than in the 
semileptonic mode. DELPHI uses $\ell\phi$ correlations from
$\bar{B}^0_s\rightarrow D^+_s\ell^-\bar{\nu}X$, $D^+_s\rightarrow\phi X$,
which has an estimated purity of $50\pm15$\%, and inclusive $D^+_s$,
which has an estimated purity of $55\pm15$\%, and significant backgrounds from
$B\rightarrow D_s D X$, and $Z^0\rightarrow c\bar{c}$, $c\rightarrow D^+_s$.
Finally, CDF reports a low statistics measurement based on an exclusive sample
from $\bar{B}^0_s\rightarrow J/\psi\phi$.

A summary of the different $\bar{B}^0_s$ lifetime measurements is
provided in Table~\ref{tab:b0s}.

\begin{table}[h] \centering
\tcaption{Summary of $\bar{B}^0_s$ lifetimes.The first error is the
statistical error, and the second error is the systematic error.}
\label{tab:b0s}
\small
\begin{tabular}{||c|c|c|c|c||}\hline\hline
Experiment & Method & Data Set & Result (psec) & Reference \\
\hline
ALEPH  & $\ell + D_s$         & 91--94 & $1.59^{+0.17}_{-0.15}\pm0.03$ &
\cite{ref:aleph_bstau}    \\
\hline
ALEPH  & $D_s+$ hadron        & 91--93 & $1.61^{+0.30+0.18}_{-0.29-0.16}$ &
\cite{ref:aleph_dshad}    \\
\hline
DELPHI & $\ell + D_s$         & 91--94 & $1.54^{+0.31}_{-0.27}\pm0.06$ &
\cite{ref:delphi_eps0559} \\
\hline
DELPHI & $D_s^+$ hadron       & 92--94 & $1.57^{+0.45+0.15}_{-0.37-0.14}$ &
\cite{ref:delphi_eps0559} \\
\hline
DELPHI & $\ell + \phi$        & 93--94 & $1.45^{+0.20+0.32}_{-0.23-0.16}$ &
\cite{ref:delphi_eps0559} \\
\hline
DELPHI & inclusive $D_s+$     & 92--94 & $1.61^{+0.34+0.18}_{-0.29-0.13}$ &
\cite{ref:delphi_eps0559} \\
\hline
OPAL   & $\ell + D_s$         & 90--94 & $1.54^{+0.25}_{-0.21}\pm0.06$ &
\cite{ref:opal_bstau}    \\
\hline
\multicolumn{3}{||c|}{LEP Average}      & $1.57\pm0.11$                &
\cite{ref:hgm_private}    \\
\hline
CDF    & $\ell + D_s$         & Run 1a & $1.42^{+0.27}_{-0.23}\pm0.11$ &
\cite{ref:cdf_bstau}      \\
\hline
CDF    & excl.~($J/\psi\phi$) & Run 1a & $1.74^{+1.08}_{-0.69}\pm0.07$ & 
\cite{ref:cdf_bstau}      \\
\hline\hline
\multicolumn{3}{||c|}{World Average}      & 
\multicolumn{2}{|c||}{$1.55^{+0.11}_{-0.10}$} \\
\hline\hline
\end{tabular} 
\end{table}

\subsection{$b$-baryons}

Two experiments have reported new measurements of the mass of the $\Lambda_b$.
Based on an analysis of $3.3\times10^6$ hadronic $Z^0$ decays
(1991--1994 data), ALEPH\cite{ref:aleph_eps0401} has
isolated four $\Lambda^0_b\rightarrow\Lambda^+_c\pi^-$ candidates
(a $2.5\sigma$ signal), which give $m(\Lambda^0_b)=5621\pm17\pm15$~MeV/$c^2$,
where the first error is statistical, and the second error is systematic.
DELPHI reports\cite{ref:delphi_eps0561}
$m(\Lambda^0_b)=5656\pm22\pm6$~MeV/$c^2$ based on three
$\Lambda^+_c\pi^-$ candidates from an analysis of $3\times10^6$ hadronic
$Z^0$ decays (1991--1994 data). Combining these new
measurements\footnote{DELPHI has a candidate for
$\Lambda^0_b\rightarrow D^0 p \pi^-$, which has not been
included in the average.}\ \ with the current world
average\cite{ref:pdg} yields $m(\Lambda^0_b)=5639\pm15$~MeV/$c^2$.
One potential problem with the $\Lambda^+_c\pi^-$ signature is that it is
difficult to distinguish this final state from final states such as
$\Lambda^+_c\rho^-$, followed by $\rho^-\rightarrow\pi^-\pi^0$,
and the $\pi^0$ is not observed. This incomplete reconstruction could
cause a systematic underestimation of $m(\Lambda^0_b)$.
Since the time of this conference, CDF has
reported\cite{ref:cdf_lbmass}
a preliminary measurement $m(\Lambda^0_b)=5623\pm5\pm4$~MeV/$c^2$
based on the observation of 38 candidates of the decay
$\Lambda^0_b\rightarrow J/\psi\Lambda$ from 115~pb$^{-1}$ (Run 1a and 1b).
The background estimate is 18; the signal significance is $3\sigma$.
The measured mass is carefully calibrated against known signals such as
$\bar{B}^0\rightarrow J/\psi K^0_s$, yielding the result
$m(\Lambda^0_b)-m(\bar{B}^0)=342\pm6$~MeV/$c^2$.

Searches for the $\Lambda^0_b$ were reported as well. Using the same data
sample listed above, DELPHI has searched for the decay
$\Lambda^0_b\rightarrow J/\psi\Lambda$ and reports\cite{ref:delphi_eps0561}
a limit
$f(b\rightarrow\Lambda^0_b)\cdot B(\Lambda^0_b\rightarrow J/\psi\Lambda)<
7\times10^{-4}$ at 90\% C.L., where $f(b\rightarrow\Lambda^0_b)$ is the
fraction of $b$~quarks that fragment into a $\Lambda^0_b$.
OPAL reports\cite{ref:opal_eps0262} the following limits based on
$1.9\times10^6$ hadronic $Z^0$ decays (1990--1993 data):
$f(b\rightarrow\Lambda^0_b)\cdot B(\Lambda^0_b\rightarrow J/\psi\Lambda)<
3.4\times10^{-4}$ at 90\% C.L. and
$f(b\rightarrow\Lambda^0_b)\cdot B(\Lambda^0_b\rightarrow\Lambda_c^+\pi^-)<
2.0\times10^{-3}$ at 90\% C.L.
%In addition to searching for the $J/\psi\Lambda$ final state, CDF has
%searched\cite{ref:cdf_lbsearch} for the $J/\psi\Lambda(1520)$
%in 70~pb$^{-1}$ of data yielding the limit
%$f(b\rightarrow\Lambda^0_b)\cdot B(\Lambda^0_b\rightarrow J/\psi\Lambda(1520))<
%2.8\times10^{-4}$ at 90\% C.L., when $p_t(\Lambda(1520))>2$ GeV/$c$.

Two principal signatures have been used to measure $\tau(\Lambda^0_b)$,
both of which are based on the semileptonic decay
$\Lambda^0_b\rightarrow\Lambda^+_c\ell^-\bar{\nu} X$.
The first signature is opposite-sign $\ell^{\pm}\Lambda^{\mp}_c$ pairs,
where the $\Lambda_c^+$ is fully reconstructed ({\it e.g.}~using final
states like $Kp\pi$ and $\Lambda\pi\pi\pi$). This signature is probably
the most unambiguous for the $\Lambda^0_b$, barring exclusive reconstruction,
but yields fairly low statistics. Conceptually it is very similar to
the other $\ell+D$ analyses above: {\it e.g.}~like-sign pairs
$\ell^{\pm}\Lambda^{\pm}_c$ and the $\Lambda^+_c$ sideband can
be used to study the backgrounds, and kinematic cuts on the lepton can
be used to suppress backgrounds.
In the second signature, the $\Lambda^+_c$ is not fully reconstructed,
instead the inclusive decay $\Lambda^+_c\rightarrow\Lambda X$ is used.
The signal is $\ell^-\Lambda$ pairs (and not $\ell^-\bar{\Lambda}$).
This signature was used by the LEP experiments to establish the existence
of $b$~baryons. The main background is leptons from semileptonic
$B$~meson decay accompanied by $\Lambda$~baryons produced in the fragmentation
of the $B$. The sample has much higher statistics than the
$\ell^{\pm}\Lambda^{\mp}_c$ sample, but suffers from two problems:
\begin{enumerate}
\item{
$b$~baryons other than the $\Lambda_b$ can contribute:
$\Xi_b\rightarrow\Xi_c\ell^-\bar{\nu}$, $\Xi_c\rightarrow\Lambda X$
yields $\ell^-\Lambda$ pairs.}
\item{
Determining the $\Lambda^0_b$ decay point is difficult, since the $\Lambda$
is very long lived and must be extrapolated a long distance back to the
lepton. Alternatively, the impact parameter of the lepton can be used
to measure the lifetime.}
\end{enumerate}

%The uncertainty of the polarization of the $\Lambda_b$
%contributes to the systematic error of a lifetime based on impact parameter.
The lifetime determined from the impact parameter distribution has a larger
systematic uncertainty than the lifetime determined using the decay length.
On the $Z^0$~resonance, $b$~quarks are produced with $-94$\% polarization.
The weakly decaying $B$~mesons are spin-$0$ and are unpolarized.
The $\Lambda_b$, however, is spin-$\frac{1}{2}$ and may be polarized.
Using a sample of $\ell^-(\Lambda\pi^+)$ pairs, ALEPH
measures\cite{ref:aleph_lbpol} the $\Lambda_b$ polarization
from the distribution of
$\langle E_{\ell}\rangle/\langle E_{\nu}\rangle$\cite{ref:bonvicini_randall},
where $\langle E_{\ell}\rangle$ is the mean energy of the leptons in the
sample and $\langle E_{\nu}\rangle$ is the mean energy of the neutrino.
The measured polarization is
$P(\Lambda^0_b)=-0.23^{+0.24+0.08}_{-0.20-0.07}$; the initial $b$~quark
polarization can be diminished due to gluon radiation and cascade decays
such as $\Sigma_b\rightarrow\Lambda_b\pi$.

OPAL measures the lifetime in the $\ell^-\Lambda$ sample using both
the decay length and the lepton impact parameter. The lifetime measured
with the lepton impact parameter is corrected by $+0.065\pm0.065$~psec
to account for possible $b$~baryon polarization, the corresponding correction
using the decay length is $+0.03\pm0.03$~psec, a factor two less.

A summary of lifetime measurements based on the $\ell^{-}\Lambda^{+}_c$
and $\ell^-\Lambda$ signatures is presented in Table~\ref{tab:bbary}.

Three of the LEP experiments report searches for the strange $b$~baryon
$\Xi_b$, which is either $\Xi^-_b$ ($bsd$) or $\Xi^0_b$ ($bsu$).
The signature is $\ell^{\pm}\Xi^{\pm}_s$~pairs from the decay sequence
$\Xi_b\rightarrow\ell^-\bar{\nu}\Xi_c X$, $\Xi_c\rightarrow\Xi^-_s X$,
$\Xi^-_s\rightarrow\Lambda\pi^-$. The physics backgrounds are
(1) $\Lambda^0_b\rightarrow\ell^-\bar{\nu}\Lambda^+_c X$,
$\Lambda^+_c\rightarrow\Xi^- K^+ \pi^+$, and
(2) $B$~meson decays such as
$B\rightarrow\Xi_c\Lambda^-_c X$, $\Xi_c\rightarrow\Xi^- X$ and
$\Lambda^-_c\rightarrow\ell^- X$. Both these backgrounds are expected
to be small, but four body semileptonic decays of the $\Lambda_b$
could also contribute to the signature and have not been estimated.

ALEPH, DELPHI, and OPAL all report excesses of same-sign
$\ell^{\pm}\Xi^{\pm}_s$~pairs over opposite-sign pairs.
ALEPH and DELPHI use this excess to calculate a signal and
determine a lifetime. OPAL reports a limit on the production
of $\Xi_b$, since there are sources of $\ell^{\pm}\Xi^{\pm}_s$~pairs
of unknown magnitude from $\Lambda_b$ decay. The results are
summarized in Table~\ref{tab:xib}.

\begin{table}[h] \centering
\tcaption{Summary of $b$ baryon lifetimes. The first error is the
statistical error, and the second error is the systematic error.}
\label{tab:bbary}
\small
\begin{tabular}{||c|c|c|c|c||}\hline\hline
Experiment & Method & Data Set & Result (psec) & Reference \\
\hline
ALEPH  & $\ell + \Lambda_c$   & 91--94 & $1.24^{+0.15}_{-0.14}\pm0.05$ &
\cite{ref:aleph_eps0753}  \\
\hline
DELPHI & $\ell + \Lambda_c$   & 91--94 & $1.26^{+0.26+0.03}_{-0.22-0.05}$ &
\cite{ref:delphi_eps0564} \\
\hline
OPAL   & $\ell + \Lambda_c$   & 90--94 & $1.14^{+0.22}_{-0.19}\pm0.07$ &
\cite{ref:opal_lblctau}   \\
\hline
ALEPH  & $\ell + \Lambda$     & 91--94 & $1.21\pm0.09\pm0.07$          &
\cite{ref:aleph_eps0753}  \\
\hline
DELPHI & $\ell + \Lambda$     & 91--94 & $1.10^{+0.16+0.05}_{-0.14-0.08}$ &
\cite{ref:delphi_eps0564} \\
\hline
DELPHI & $\mu + p$            & 92     & $1.27^{+0.35}_{-0.29}\pm0.09$ &
\cite{ref:delphi_lblp}    \\
\hline
OPAL   & $\ell + \Lambda$     & 90--94 & $1.16\pm0.11\pm0.06$          &
\cite{ref:opal_lbllamtau} \\
\hline
\multicolumn{3}{||c|}{LEP Average}      & $1.18\pm0.07$                &
\cite{ref:hgm}            \\
\hline
ALEPH  & $\ell + \Xi$         & 91--94 & $1.25^{+0.55}_{-0.35}\pm0.20$ &
\cite{ref:aleph_eps0406}  \\
\hline
DELPHI & $\ell + \Xi$         & 91--93 & $1.50^{+0.7}_{-0.4}\pm0.30$   &
\cite{ref:delphi_strbbar} \\
\hline
\multicolumn{3}{||c|}{Average $\ell + \Xi$} & $1.36^{+0.44}_{-0.33}$   &
\cite{ref:hgm}            \\
\hline\hline
\end{tabular}
\end{table}

\begin{table}[h] \centering
\tcaption{Summary of results on the strange $b$ baryon $\Xi_b$. When
two errors are listed, the first error is the statistical error,
and the second error is the systematic error. The DELPHI lifetime
result is based on a subset of the data sample: $1.7\times10^6 Z^0$
(91--93), which has 10 candidates (7 same-sign and 3 opposite-sign).}
\label{tab:xib}
\small
\begin{tabular}{||c|c|c|c||}\hline\hline
Experiment                        & ALEPH & DELPHI & OPAL \\
\hline
\# hadronic $Z^0$                 & $3.45\times10^6$ (90--94)  &
                                    $3.06\times10^6$ (91--94)  &
                                    $3.61\times10^6$ (90--94)   \\
\hline
Excess of $\ell^{\pm}\Xi^{\pm}_s$ & $22.5\pm5.7$ & $14.0\pm3.7$ & $29\pm14$ \\
over $\ell^{\pm}\Xi^{\mp}_s$      &            &            &         \\
\hline
$f(b\rightarrow\Xi_b)\times$      & $(5.3\pm1.3\pm0.7)$ &
                                    $(6.6\pm1.7\pm1.0)$ &
                                    $<5.1\times10^{-4}$  \\
$B(\Xi_b\rightarrow\Xi\ell\nu X)$ & $\times10^{-4}$     &
                                    $\times10^{-4}$     &
                                    at 95\% C.L.         \\
\hline
$\tau$ (psec)                     & $1.25^{+0.55}_{-0.35}\pm0.20$  &
                                    $1.50^{+0.7}_{-0.4}\pm0.30$    &
                                    --                               \\
\hline
Reference                         & \cite{ref:aleph_eps0406}       &
                                    \cite{ref:delphi_eps0564}      &
                                    \cite{ref:opal_eps0272}          \\ 
\hline\hline
\end{tabular}
\end{table}

\subsection{Search for $B_c$}

The $B^-_c$~meson is a bound state of a $b$~quark and a $\bar{c}$~quark
($b\bar{c}$) that decays weakly and therefore is very narrow.
Reliable calculations of the mass and decay of this bound state
exist. There are 15 expected states below $BD$~threshold.
Some of the predicted properties\cite{ref:quigg_bc} are
$m(B^-_c)=6256\pm20$~MeV/$c^2$,
$\tau(B^-_c)=1.35\pm0.15$~psec\footnote{Some lifetime calculations predict
significantly shorter lifetimes.}\ \
$B(B^-_c\rightarrow J/\psi X)\sim10$\%.
At the Tevatron, the production rate is expected to be $~\sim10^{-3}$ of
the production rate of the lighter $B$~hadrons; at the $Z^0$~resonance,
a few hundred $B^-_c$ are expected for every $10^6$ hadronic $Z^0$ decays.

%The search for the $B^-_c$ has been made more exciting by a related
%result\cite{ref:opal_ups}
%submitted by the OPAL. They have observed inclusive $\Upsilon$ production
%at the $Z^0$. Starting with a sample of $3.7\times10^6$ hadronic $Z^0$ decays,
%they find 6 candidates for the process
%$Z^0\rightarrow\Upsilon X$, $\Upsilon\rightarrow\mu^+\mu^-$ or
%$\Upsilon\rightarrow e^+e^-$. The measured branching fraction is
%$B(Z^0\rightarrow\Upsilon X)=1.2^{+0.9+0.2+1.7}_{-0.6-0.2-0.4}$,
%where the first error is statistical, the second is the experimental systematic
%error, and the third error is the systematic error due to model dependence.
%This branching fraction is about an order of magnitude higher than expected.
%Three possible production mechanisms are shown in Figure~\ref{fig:bc}.
%If production mechanism (a) is the main source of $\Upsilon$s, then replacing
%$g\rightarrow b\bar{b}$ with $g\rightarrow c\bar{c}$ gives a production
%mechanism for $B^-_c$.

Related to the search for the $B^-_c$ is the observation\cite{ref:opal_ups}
of $\Upsilon$ production at the $Z^0$ by OPAL.
Starting with a sample of $3.7\times10^6$ hadronic $Z^0$ decays,
they find 8 candidates for the process
$Z^0\rightarrow\Upsilon X$, $\Upsilon\rightarrow\mu^+\mu^-$ or
$\Upsilon\rightarrow e^+e^-$, where $\Upsilon$ means either
$\Upsilon(1S)$, $\Upsilon(2S)$, or $\Upsilon(3S)$, which OPAL can not
distinguish experimentally. The estimated background is $1.6\pm0.3$.
The measured branching fraction is
$B(Z^0\rightarrow\Upsilon X)=(1.0\pm0.4\pm0.1\pm0.2)\times 10^{-4}$,
where the first error is statistical, the second is the experimental systematic
error, and the third error is the systematic error due to uncertainties
in the production mechanism. The observed rate is an order of magnitude larger
than the rate expected from so-called colour-singlet models\cite{ref:opal_ups},
which are dominated by the production of $\Upsilon$'s via $b$~quark
fragmentation: $Z^0\rightarrow b\bar{b}, b\rightarrow\Upsilon$.
This same process should lead to $B^-_c$ formation. The discrepancy between
the expected and observed rate could originate from (1) a statistical
fluctuation, (2) additional production mechanisms such as those predicted
by so-called colour-octet models\cite{ref:opal_ups}, or (3) an underestimate
of the production via $b$~quark fragmentation, which could indicate that the
$B^-_c$ production may be underestimated as well.

%
%\begin{figure}
%\begin{picture}(100,100)
%\put (0,0){\line(1,0){424}}
%\put (0,100){\line(1,0){424}}
%\put (25,50){Three production mechanisms for $Z^0\rightarrow\Upsilon X$}
%\end{picture}
%\fcaption{Three possible production mechanisms for $Z^0\rightarrow\Upsilon X$.
%Mechanism (a) is predicted to be the largest contribution. Replacing
%$g\rightarrow b\bar{b}$ with $g\rightarrow c\bar{c}$ yields a mechanism for
%producing $B^-_c$.}
%\label{fig:bc}
%\end{figure}
%
The $B^-_c$ has a clean signature, based on the spectator decay in which
the $b$~quark decays to a $c$~quark and a virtual $W^*$;
the $c$ and spectator $\bar{c}$ form a $J/\psi$ and the $W^*$ becomes
$\ell^-\bar{\nu}$ or a $\pi^-$. The predicted branching fractions
(including the $J/\psi$ branching fractions) are
$B(B^-_c\rightarrow J/\psi\pi^-)\cdot B(J/\psi\rightarrow\ell^+\ell^-)
\sim3\times10^{-4}$, and
$B(B^-_c\rightarrow J/\psi\ell^-\bar{\nu})\cdot B(J/\psi\rightarrow\ell^+\ell^-)
\sim10^{-3}$.

Two experiments have reported\footnote{Following the symposium, OPAL has
submitted a search\cite{ref:opal_jpsi} for the $B^-_c$ for
publication.}\ \ searches.
Starting with a sample of 600 $J/\psi$ candidates from
$3.1\times10^6$ hadronic $Z^0$ decays (1991--1994),
ALEPH finds\cite{ref:aleph_eps0407}
no candidates in the final state $J/\psi\pi$,
one candidate in the final state $J/\psi\mu\nu$, and
one candidate in the final state $J/\psi e\nu$.
The expected backgrounds are 0.32, 0.17, and 0.13, respectively.
The probability for the observed two events to come from the expected
background is 4\%, and they report the following limits at 90\% C.L.:
$$\frac{B(Z\rightarrow B_c X)}{B(Z\rightarrow q\bar{q})}
\cdot B(B^-_c\rightarrow J/\psi\pi^-)<4\times10^{-5},$$
$$\frac{B(Z\rightarrow B_c X)}{B(Z\rightarrow q\bar{q})}
\cdot B(B^-_c\rightarrow J/\psi\ell^-\bar{\nu})<7\times10^{-5}.$$
CDF searches\cite{ref:cdf_bc}
for a peak in the invariant mass distribution of
$J/\psi\pi^-$ candidates between 6.1 GeV/$c^2$ and 6.4 GeV/$c^2$.
No signal is observed, and they set a limit
%(see Figure~\ref{fig:cdf_bc})
on the product of the production cross-section times the branching fraction,
$\sigma(p\bar{p}\rightarrow B^-_c X)\cdot B(B^-_c\rightarrow J/\psi\pi^-)$,
normalized to their observed $B^+\rightarrow J/\psi K^-$ signal.
The limit is 0.12 for $\tau(B^-_c)=0.17$ psec and decreases to 0.068 for
$\tau(B^-_c)=1.6$ psec.

\section{Results on $B^*$ and $B^{**}$}

In Heavy Quark Effective Theory (HQET), the heavy and light quarks decouple
as the heavy quark mass increases. The spin-angular momentum of the heavy quark,
$\vec{s}_Q$ and the total angular momentum of the light quark,
$\vec{j}_q=\vec{l}+\vec{s}_q$, are conserved separately.
A $Q\bar{q}$ bound state has total angular momentum
$\vec{J}=\vec{s}_Q\oplus\vec{j}_q$, so for each value of $j_q$ there is
a doublet. In the case of $B$~mesons, the doublet of particles that
corresponds to the case in which the light quark has zero angular
momentum ($l=0$) is the familiar $B$ ($J^P=0^-$) and $B^*$ ($J^P=1^-$)
mesons. For $l=1$, there are two values of $j_q$ ($\frac{1}{2}$ and
$\frac{3}{2}$), and two corresponding doublets. These four p-wave mesons
are collectively referred to as the $B^{**}$'s.
Their expected properties\cite{ref:ehq} are summarized in
Table~\ref{tab:bpwave}.

\begin{table}[h] \centering
\tcaption{The expected p-wave $B$~meson ($b\bar{d}$ and $b\bar{u}$) states
and their predicted\cite{ref:ehq} masses, widths, and decay modes.
In the final column (Decay Modes), the $L$ refers to the orbital angular
momentum of the $B\pi$ system. Angular momentum conservation and parity
require $L=0$ or $L=2$. For the predicted $B^{**}$ masses, $B\rho$ decays
are suppressed by phase-space.}
\label{tab:bpwave}
\small
\begin{tabular}{||c|c|c|c|c||}\hline\hline
State & $J^P(j_q)$ & Mass (MeV/$c^2$) & Width (MeV/$c^2$) & Decay Modes \\
\hline
$B^*_2$ & $2^+(\frac{3}{2})$ & 5771 & 25 & $(B^*\pi)_{L=2},(B\pi)_{L=2}$ \\
\hline
$B_1$   & $1^+(\frac{3}{2})$ & 5759 & 21 & $(B^*\pi)_{L=2}$ \\
\hline
$B_1$   & $1^+(\frac{1}{2})$ & $\sim5670$ & broad & $(B^*\pi)_{L=0}$ \\
\hline
$B^*_0$ & $0^+(\frac{1}{2})$ & $\sim5670$ & broad & $(B\pi)_{L=0}$ \\
\hline\hline
\end{tabular}
\end{table}

Observation of a new state is always interesting in itself, and the
observed properties of the $B^{**}$'s provide a test of predictions
of HQET. The real motivation for searching for $B^{**}$'s, however, is that
they may provide an effective $b$ flavour tag for CP-violation experiments.
For example, a $B^{*-}_2$ ($b\bar{u}$) can decay to a $\bar{B}^0\pi^-$;
the negative charge of the pion signals the flavour of the $\bar{B}^0$
at production. Actually the idea of exploiting the pions
from the $B^{**}$ resonances as a flavour tag was preceeded by the
idea\cite{ref:gnr} of using associated fragmentation particles as a
flavour tag. For example, if a $b$~quark is produced, and a $d\bar{d}$~pair
materializes out of the vacuum, then the $b$ and the $\bar{d}$ may form
a $\bar{B}^0$~meson, and the left over $d$~quark could combine with a
$\bar{u}$~quark to form a $\pi^-$. As in the above example, the charge of the
pion signals that a $\bar{B}^0$ was produced. The narrow $B^{**}$
resonances, however, potentially provide much better signal-to-noise and,
therefore, a cleaner flavour tag. Results on flavour tagging from ALEPH
and OPAL using these methods are presented later in this section.

Results on excited $B_s$ states and $b$~baryon states were also submitted
to this conference. The expected properties of these particles are summarized
in Tables~\ref{tab:bspwave} and~\ref{tab:sigmab}. A significant production
of $B^{**}_s$ actually reduces the production of weakly decaying $B^0_s$.
Furthermore, since $B^{**}_s\rightarrow B^0 \bar{K}^0$, and the $\bar{K}^0$
is neutral, $B^{**}_s$ production does not provide a flavour tag for $B^0$.
A significant production of $\Sigma^{(*)}_b$ reduces the polarization of
weakly decaying $\Lambda^0_b$.

\begin{table}[h] \centering
\tcaption{The expected p-wave $B_s$~meson ($b\bar{s}$) states
and their predicted\cite{ref:ehq} masses, widths, and decay modes.
In the final column (Decay Modes), ``$B$'' means $b\bar{u}$ or
$b\bar{d}$. The widths of the $j_q=\frac{1}{2}$ doublet are hard
to predict since the $BK$ decays may be phase-space
suppressed. $B_s\pi$ decays are forbidden by isospin.}
\label{tab:bspwave}
\small
\begin{tabular}{||c|c|c|c|c||}\hline\hline
State & $J^P(j_q)$ & Mass (MeV/$c^2$) & Width (MeV/$c^2$) & Decay Modes \\
\hline
$B^*_{s2}$ & $2^+(\frac{3}{2})$ & 5849 & 1 & $(B^* K)_{L=2},(BK)_{L=2}$ \\
\hline
$B_{s1}$   & $1^+(\frac{3}{2})$ & 5861 & 4 & $(B^*K)_{L=2}$ \\
\hline
$B_{s1}$   & $1^+(\frac{1}{2})$ & $\sim5750$ & ? & $(B^* K)_{L=0}$ or 
                                                   $B_s\gamma$ \\
\hline
$B^*_{s0}$ & $0^+(\frac{1}{2})$ & $\sim5750$ & ? & $(BK)_{L=0}$ or
                                                   $B_s\gamma$ \\
\hline\hline
\end{tabular}
\end{table}

\begin{table}[h] \centering
\tcaption{The reported excited $b$~baryons and their expected
properties\cite{ref:roncaglia}. The total spin angular momentum of the two
light quarks is denoted by $s_{qq}$, and $\vec{J}=\vec{s}_b\oplus\vec{s}_{qq}$.}
\label{tab:sigmab}
\small
\begin{tabular}{||c|c|c|c||}\hline\hline
State & $J^P(s_{qq})$ & $M(\Sigma_b)-M(\Lambda_b)$ (MeV/$c^2$) & Decay Modes \\
\hline
$\Lambda^0_b$  & $\frac{1}{2}^+(0)$ & --         & weak\\
\hline
$\Sigma_b$     & $\frac{1}{2}^+(1)$ & $200\pm20$ & $\Lambda^0_b\pi$ \\
\hline
$\Sigma^{*}_b$ & $\frac{3}{2}^+(1)$ & $230\pm20$ & $\Lambda^0_b\pi$ \\
\hline\hline
\end{tabular}
\end{table}

Results on $B^{*}$ and $B^{**}$ were reported by the LEP collaborations.
There are two approaches (1) {\it inclusive $B$ reconstruction}
(results from ALEPH, DELPHI, L3, and OPAL), which yields large samples
($\sim10^5$) of $B$'s; and (2) {\it exclusive $B$ reconstruction}
(results from ALEPH), which is cleaner, but has much lower statistics
($\sim500$ $B$'s).

\subsection{Inclusive $B$ Hadron Reconstruction}

Inclusive $B$~hadron reconstruction is possible because of the long
$B$~hadron lifetime, the large $B$~hadron mass, and the hard
$b$~quark fragmentation. There have been three approaches.
The first published method\cite{ref:l3_bstar}
of inclusive $B$~hadron reconstruction was from
the L3 collaboration. Their approach was used to reconstruct the $B^*$ and is
discussed in more detail in that section.
The second method\cite{ref:opal_bdstar} was developed
by OPAL. It distinguishes between charged~$B$'s and neutral~$B$'s,
and as we shall see, this charge assignment makes it possible to
determine background rates and flavour tagging efficiencies from the data.
A third method\cite{ref:delphi_bstar} was developed by DELPHI,
and is also used by ALEPH\cite{ref:aleph_binc}.
This method has higher efficiency than the OPAL method, but does
not determine the $B$~hadron charge. Monte Carlo models are used to
determine the background.

The method of DELPHI and ALEPH proceeds as follows.
First $Z^0\rightarrow b\bar{b}$ events are selected using track impact
parameters. A purity of 80--90\% is achieved with good efficiency.
Next charged tracks and neutral calorimeter clusters are
combined based on their rapidity $y$ and impact parameter $\delta$.
The rapidity is defined with respect to the thrust axis of the event
or a jet axis.
Charged tracks are assigned the pion mass, and neutral clusters are
treated as photons. The large $B$ mass and hard $B$ fragmentation gives
the decay products from $B$~hadrons a harder rapidity distribution than
fragmentation products and particles not coming from $B$ decay.
The long $B$ hadron lifetime gives the charged particles from $B$~hadron
decay larger impact parameters (with respect to the interaction point)
on average. To reduce the fraction of poorly measured $B$'s, requirements
are made on the reconstructed $B$~mass, and the reconstructed energy or
momentum. The raw $B$ energy is corrected to account for undetected
neutrinos, detector inefficiencies, and incorrect particle mass
assignments. This correction depends on the reconstructed $B$~mass
and the observed energy in the hemisphere.

\begin{figure}[h] \centering
\mbox{}
\epsfysize 10cm
\epsffile{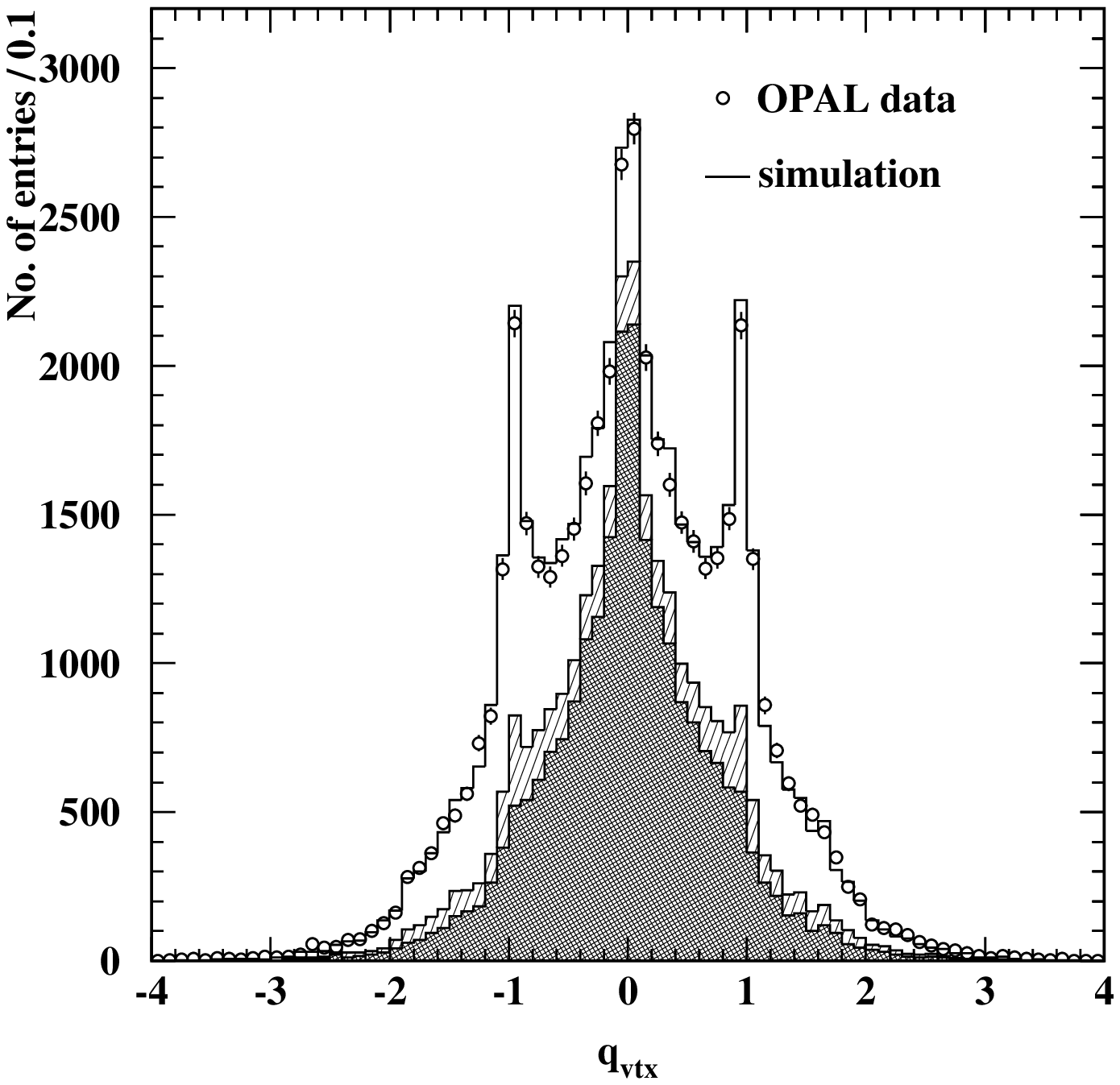}
\fcaption{The distribution of $Q_{vtx}$ observed in the OPAL data (points)
compared to the expectations from Monte Carlo simulation (histogram).
The solid region represents the expected contribution from neutral
$B$~hadrons, the hatched region shows the expected contribution from
sources other than $B$~hadrons, and the remaining contribution is from
$B^{\pm}$. The requirement $Q_{vtx}>0.6$ increases the overall fraction
of $B^{\pm}$ from $(40\pm3)$\% to $(54\pm2)$\%.}
\label{fig:opal_qvtx}
\end{figure}

The OPAL approach is to select hadronic $Z^0$ decays and then find
the jets in these events using a cone algorithm. Next they search
for secondary vertices in the two highest energy jets.
If a vertex is found, they assign a vertex charge $Q_{vtx}$,
which is defined as follows:
$$Q_{vtx}=\sum_{i=tracks} w_i q_i,$$
where $w_i$ is a weight that is larger if the track is consistent with
coming from the secondary vertex and smaller if the particle is consistent
with coming from the interaction point. The $Q_{vtx}$ distribution
observed in the data is compared with the predicted distribution from
Monte Carlo simulation in Figure~\ref{fig:opal_qvtx}.
%$Q_{vtx}$ is a continuous quantity, not a discrete quantity.
The momentum of the vertex is also calculated
using these weights, and neutral energy is added to the charged momentum
to get the total $B$~energy. The $B$~energy resolution is improved
by a factor of two by taking advantage of the known center-of-mass
energy of the collision.

The method of DELPHI and ALEPH and the method of OPAL achieve similar
resolutions. The typical $B$~hadron energy resolution is
$\sigma_E/E=7-9$\%, and the typical angular resolutions are
$\sigma_{\phi}=10-15$~mrad and $\sigma_{\theta}=15$~mrad, where
$\theta$ and $\phi$ are the polar and azimuthal angles, respectively.
These resolutions are for the core of the distributions; there
are significant non-Gaussian tails. 
%For their $B^*$ ($B^{**}$) search,
%ALEPH reconstructs 460\thinspace000 (90\thinspace000)
%$B$'s in 3~million hadronic $Z^0$~decays and OPAL reconstructs
%80\thinspace000 $B$'s in $3.5$ million hadronic $Z^0$~decays.
For their $B^*$ ($B^{**}$) search, ALEPH reconstructs
460\thinspace000 (90\thinspace000) inclusive $B$~hadrons with a purity of
$94\pm2$\% ($98.5\pm1.5$\%) in a sample of 3~million hadronic $Z^0$~decays.
DELPHI achieves similar efficiencies. OPAL reconstructs
80\thinspace000 $B$~hadrons with a purity of $89\pm2$\% in $3.5$ million
hadronic $Z^0$~decays.

\subsection{Results on $B^*$}

The mass difference\cite{ref:pdg} between the $B$ and $B^*$ mesons is
$\Delta M(B^*-B)=46\pm0.6$~MeV/$c^2$, so the
$B^*$ can decay to a $B$ only via a photon: $B^*\rightarrow B\gamma$.
On the $Z^0$~resonance, the mean energy of this $\gamma$ is 300 MeV, and
the maximum energy is 800 MeV. In addition to reconstructing the $B$, there
is the experimental challenge of measuring these relatively low energy photons.
Instead of using their electromagnetic calorimeters, ALEPH and DELPHI rely
on their tracking systems to reconstruct the photon when it converts.
Using conversions, DELPHI (ALEPH) measures the energy of these photons
down to $E_{\gamma}=100 (200)$ MeV. To increase statistics, they also
use conversion candidates in which only one leg of the conversion is
reconstructed in the tracking system. They then combine these photons with
the $B$~hadrons reconstructed inclusively as described above to form
$\Delta M = M(B\gamma) - M(B)$. They see a peak at the known $B^*$--$B$
mass difference.\footnote{The analyses presented here do not distinguish
between $B^{*0}$ and $B^{*0}_s$.}

L3 measures $E_{\gamma}$ in their high resolution BGO calorimeter;
the minimum photon energy is 100 MeV. They reconstruct $B$~hadrons
starting with a sample of high~$p$, high~$p^{rel}_t$ muons.
The muon is combined with the closest jet to form the $B$~direction.
The $B$ momentum is fixed at 37 GeV/$c$ (this approach is adequate,
again, due to the hard $B$ fragmentation). This procedure results in
a $B$ purity of 84\%, a $B$ angular resolution of 35 mrad, and an
energy resolution of 20\%. The $B^*$ signal appears as an enhancement
in the distribution of $E^{\gamma}_{rest}$, which is the energy of the
photon in the rest~frame of the $B$. Since $M(B^*)-M(B)\ll M(B)$,
the recoil of the $B$ is negligible, and $E^{\gamma}_{rest}$ is a
good approximation of the mass difference. 

The results for the mass difference $\Delta M=M(B^*)-M(B)$,
the relative production, and the polarization of the $B^*$
on the $Z^0$~resonance from ALEPH, DELPHI,
and L3 are summarized in Table~\ref{tab:bstar}. 
The measurements of the mass difference are comparable in precision to
measurements made by
CUSBII\cite{ref:cusb_bstar}, $\Delta M=45.6\pm0.8$ MeV/$c^2$, and
CLEOII\cite{ref:cleo_bstar}, $\Delta M=46.2\pm0.3\pm0.8$ MeV/$c^2$
in $e^+e^-$ collisions at $\sqrt{s}=10.61-10.70$ GeV.
The relative production,
denoted $N_{B^*}/(N_{B^*}+N_{B})$ is the number of $B^*$ produced divided
by the total number of $B^*$ and $B$ produced. Based on spin~counting,
and neglecting the small mass difference $M(B^*)-M(B)$, this ratio is
expected to be 0.75. Significant production of $B^{**}$ could alter this
expectation. The equivalent ratio measured\cite{ref:aleph_dstar} for charm is
$0.51\pm0.04$. ALEPH and DELPHI have measured the fraction of longitudinally
polarized $B^*$, $\sigma_L/(\sigma_L+\sigma_T)$, using the angular distribution
of the photon in the $B^*$ rest frame. The expected value of this ratio,
based on spin~counting, is 0.33.
The measured values of $N_{B^*}/(N_{B^*}+N_{B})$ and
$\sigma_L/(\sigma_L+\sigma_T)$ agree with the expectations based
on spin~counting.

\begin{table}[h] \centering
\tcaption{Summary of results for the mass difference $\Delta M=M(B^*)-M(B)$,
the relative production, and the polarization of the $B^*$ on the
$Z^0$~resonance from ALEPH, DELPHI, and L3. The L3 experiment reports
the energy of the photon in the $B$ restframe, rather than $M(B^*)-M(B)$.
Since $M(B^*)-M(B)\ll M(B)$,
the recoil of the $B$ is negligible, and $E^{\gamma}_{rest}$ is a
good approximation of the mass difference. For all reported measurements,
the first error is statistical and the second error is systematic.}
\label{tab:bstar}
\small
\begin{tabular}{||c|c|c|c||}\hline\hline
Experiment                        & ALEPH & DELPHI & L3 \\
\hline
\# hadronic $Z^0$                 & $3.0\times10^6$ (91--94)  &
                                    $2.3\times10^6$ (91--94)  &
                                    $1.6\times10^6$ (91--93)   \\
\hline
$\Delta M$ (MeV/$c^2$)            & $45.3\pm0.35\pm0.87$ &
                                    $45.5\pm0.3 \pm0.8 $ &
                                    $46.3\pm1.9$ (stat.)  \\
\hline
$\frac{N_{B^*}}{N_{B^*}+N_{B}}$ (\%) & $77.1\pm2.6\pm7.0$ &
                                       $72  \pm3  \pm6  $ &
                                       $76  \pm8  \pm6  $  \\
\hline
$\frac{\sigma_L}{\sigma_L+\sigma_T}$ (\%) & $33\pm6\pm5$ &
                                            $32\pm4\pm3$ &
                                            --            \\
\hline
Reference                         & \cite{ref:aleph_binc}   &
                                    \cite{ref:delphi_bstar} &
                                    \cite{ref:l3_bstar}      \\ 
\hline\hline
\end{tabular}
\end{table}

\subsection{Results on $B^{**}$}

\begin{figure}[h] \centering
\mbox{}
\epsfysize 6cm
\epsffile{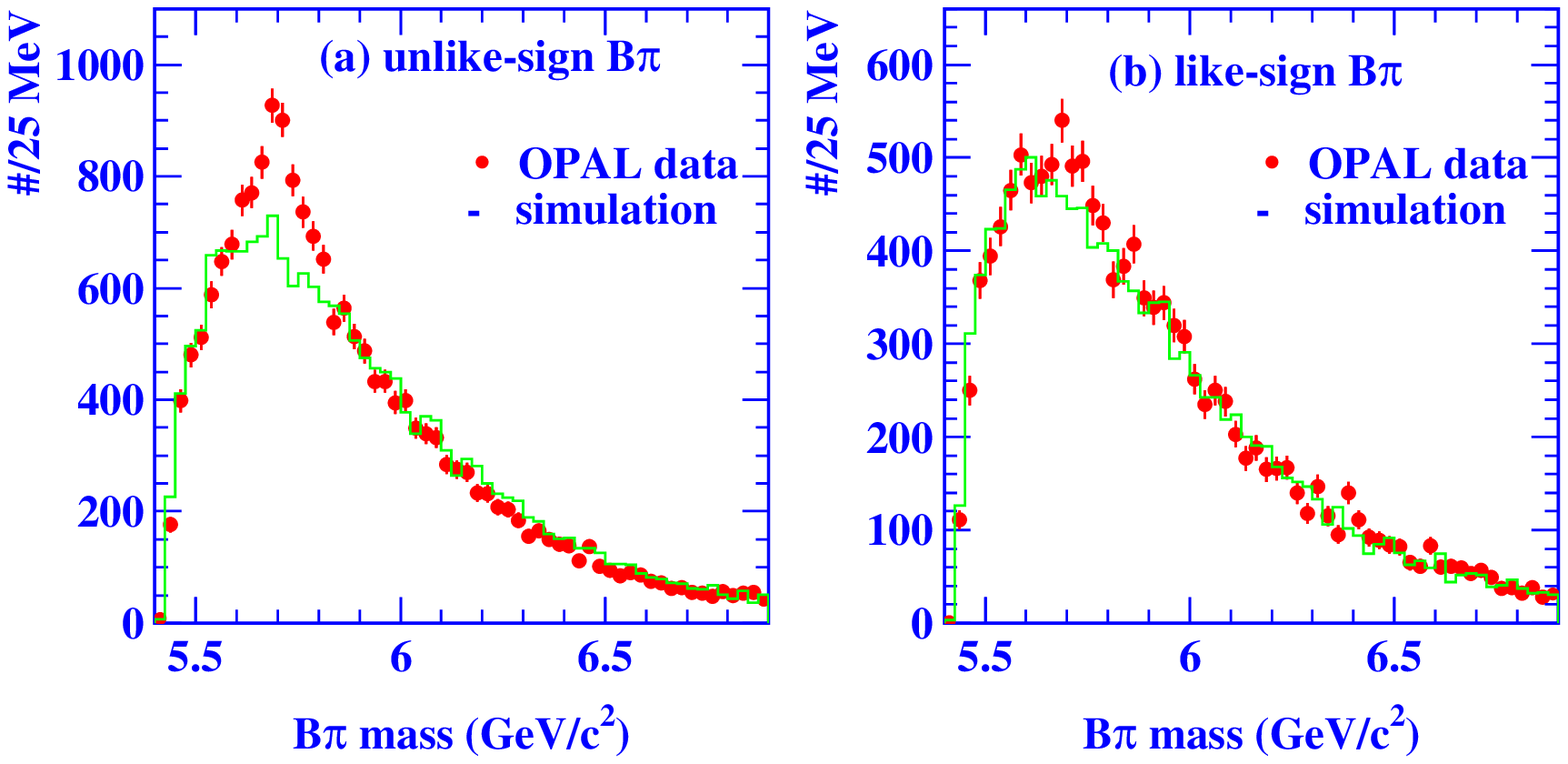}
\fcaption{The $B\pi$ mass distribution from OPAL for
(a) unlike-sign $B\pi$ combinations, which are defined by
$q_{vtx}\cdot q_{\pi}<-0.71$, and for (b) like-sign $B\pi$ combinations,
which are defined by $q_{vtx}\cdot q_{\pi}>0.49$. The points are the data,
and the histogram is the Monte Carlo simulation, which does not contain
$B^{**}$. The Monte Carlo is normalized to the same number of secondary
vertices observed in the data. The observed enhancement in the unlike-sign
data is attributed to $B^{**}$ production.}
\label{fig:opal_bdstar}
\end{figure}

\begin{figure}[h] \centering
\mbox{}
\epsfysize 10cm
\epsffile{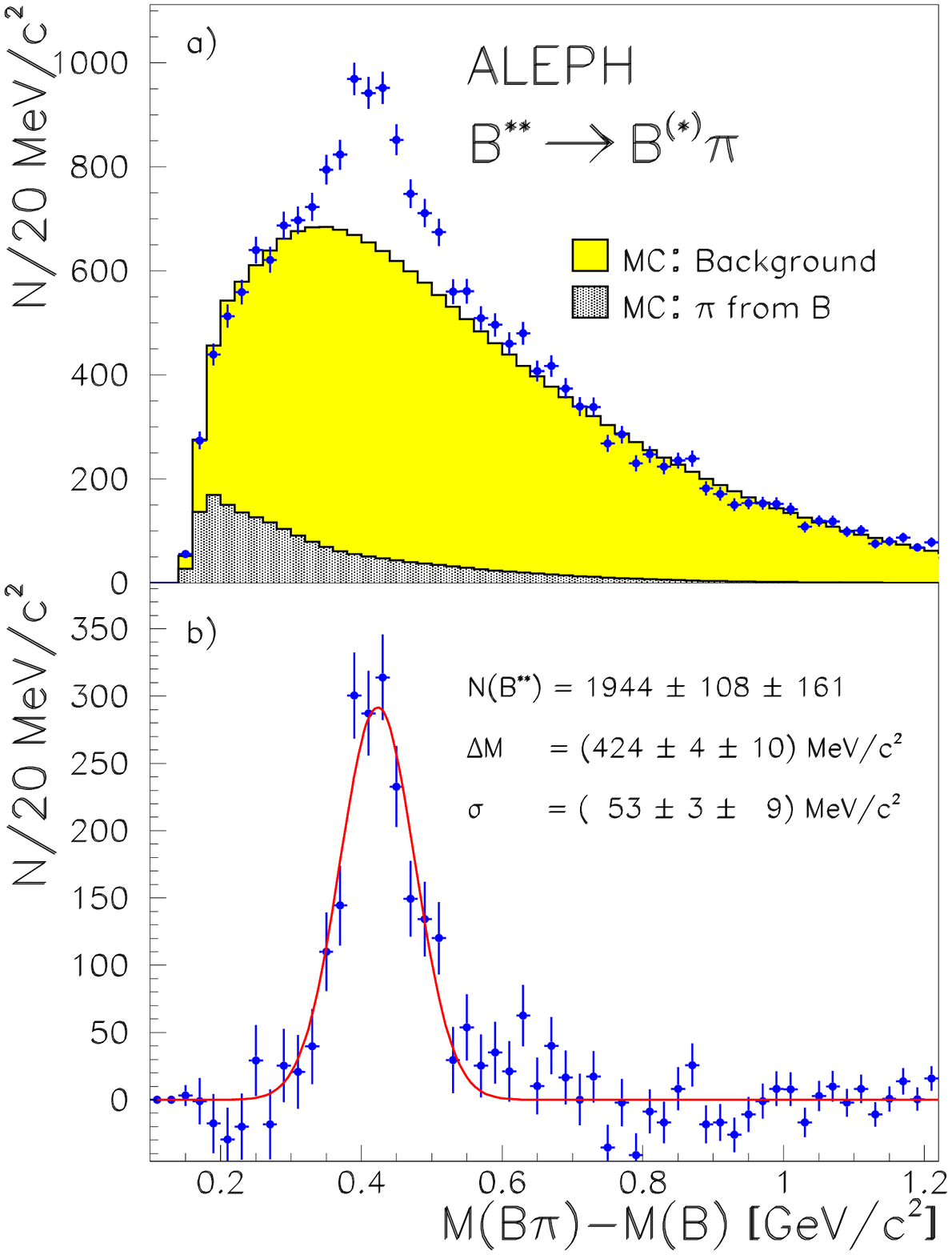}
\fcaption{(a) The $Q=M(B\pi)-M(B)$ distribution observed by ALEPH. The points
are the data and the histogram is the background predicted by the Monte Carlo,
normalized to fit the data in the sidebands,
$Q<0.25$~GeV/$c^2$ and $0.7<Q<1.2$~GeV/$c^2$.
The lower histogram shows the expected background from soft pions from
$B$ decay. The modelling of these soft pions has been
adjusted\cite{ref:aleph_binc} according to data from the $\Upsilon(4S)$.
(b) The difference between the data and the Monte Carlo. The observed
enhancement is fit to a Gaussian, resulting in the parameters shown
in the Figure and listed in Table~\ref{tab:bud_dstar}.}
\label{fig:aleph_inclb}
\end{figure}

ALEPH, DELPHI, and OPAL combine their inclusively reconstructed $B$'s with
charged pions and look for resonant structure in the $B\pi$ mass distribution
(or the equivalent $Q$ distribution, where $Q=M(B\pi)-M(B)$ or
$Q=M(B\pi)-M(B)-M(\pi)$). All three experiments require that candidate pions
are (1) consistent with coming from the primary interaction point (this
reduces combinatoric background from charged particles from $B$ decay), and 
are (2) identified as pions using $dE/dx$ measurements or RICH (DELPHI only).
Using this reconstruction procedure, decays of the type
$B^{**}\rightarrow B^*\pi$, $B^*\rightarrow B\gamma$ are shifted
down in mass (or Q) by 46 MeV/$c^2$ with respect to the decays
$B^{**}\rightarrow B\pi$; they do not cause significant broadening, however,
of the observed resonant structure.
The $B\pi$ mass distribution from OPAL is shown in Figure~\ref{fig:opal_bdstar},
and the $Q$ distribution from ALEPH is shown in Figure~\ref{fig:aleph_inclb}.
All three experiments observe resonant
structure and they fit this structure with a Gaussian or Breit-Wigner.
The results are summarized in Table~\ref{tab:bud_dstar}.
The fitted widths are broader than the experimental resolution
($\approx 40$~MeV/$c^2$) expected for a single narrow state.
The experiments also fit the observed enhancement to a variety of models
based on the expected contributions of the four p-wave mesons.
The expected p-wave mesons qualitatively describe the data, but
it is not possible to distinguish contributions from a specific state.
DELPHI examined\cite{ref:delphi_eps0563} the distribution of the helicity
angle, $\alpha_{\pi}$, defined as the angle between the momentum of the
pion candidate in the $B\pi$ rest frame and the momentum of the $B\pi$
in the laboratory. Their data are consistent with a flat helicity angle
distribution.

Table~\ref{tab:bud_dstar} also summarizes the production fraction,
$\frac{B(Z\rightarrow b\rightarrow B^{**}_{u,d})}
{B(Z\rightarrow b\rightarrow B_{u,d})}$,
which is determined assuming $B^{(*)}\pi$ decays are dominant and
that $B(B^{**}\rightarrow B^{(*)}\pi^{\pm})=
2\cdot B(B^{**}\rightarrow B^{(*)}\pi^{0})$ (isospin).
A substantial fraction of $b$~quarks fragment into p-wave mesons.

\begin{table}[h] \centering
\tcaption{Summary of results on $B^{**}_{u,d}$. All results have been converted
to $Q=M(B\pi)-M(B)-M(\pi)$. The fitted parameters for Gaussians or
Breit-Wigners are reported\cite{ref:rvk}. The OPAL results are for
$B^{-}\pi^{+}$  combinations only. The production ratio is defined as
$\frac{B(Z\rightarrow b\rightarrow B^{**}_{u,d})}
{B(Z\rightarrow b\rightarrow B_{u,d})}$, and is determined assuming
$B^{(*)}\pi$ decays are dominant and isospin.
For all reported measurements,
the first error is statistical and the second error is systematic.
Note that the observed signal for DELPHI corresponds to the 91--93
data set.}
\label{tab:bud_dstar}
\small
\begin{tabular}{||c|c|c|c||}\hline\hline
Experiment                        & ALEPH & DELPHI & OPAL \\
\hline
\# hadronic $Z^0$                 & $3.0\times10^6$ (91--94)  &
                                    $2.3\times10^6$ (91--94)  &
                                    $3.5\times10^6$ (91--94)   \\
\hline
Signal                            & $1944\pm108\pm161$ &
                                    $2157\pm120\pm323$ &
                                    $1738\pm121\pm153$  \\
\hline
$Q$ (MeV/$c^2$)                   & $284\pm4\pm10$ &
                                    $285\pm5\pm12$ &
                                    $262\pm11$       \\
\hline
$\sigma$ (MeV/$c^2$)              & $53\pm3\pm9$ &
                                    $72\pm5\pm8$ &
                                    $60\pm8$      \\
\hline
$\Gamma$ (MeV/$c^2$)              & --         &
                                    $120\pm21$ &
                                    $116\pm24$  \\
\hline
Prod.~ratio (\%)                  & $27.9\pm1.6\pm5.9\pm3.8$ &
                                    $32.5\pm1.9\pm6.0$       &
                                    $27.0\pm1.2\pm5.3$        \\
\hline
Reference                         & \cite{ref:aleph_binc}     &
                                    \cite{ref:delphi_eps0563} &
                                    \cite{ref:opal_bdstar}     \\ 
\hline\hline
\end{tabular}
\end{table}

\begin{figure}[h] \centering
\mbox{}
\epsfysize 10cm
\epsffile{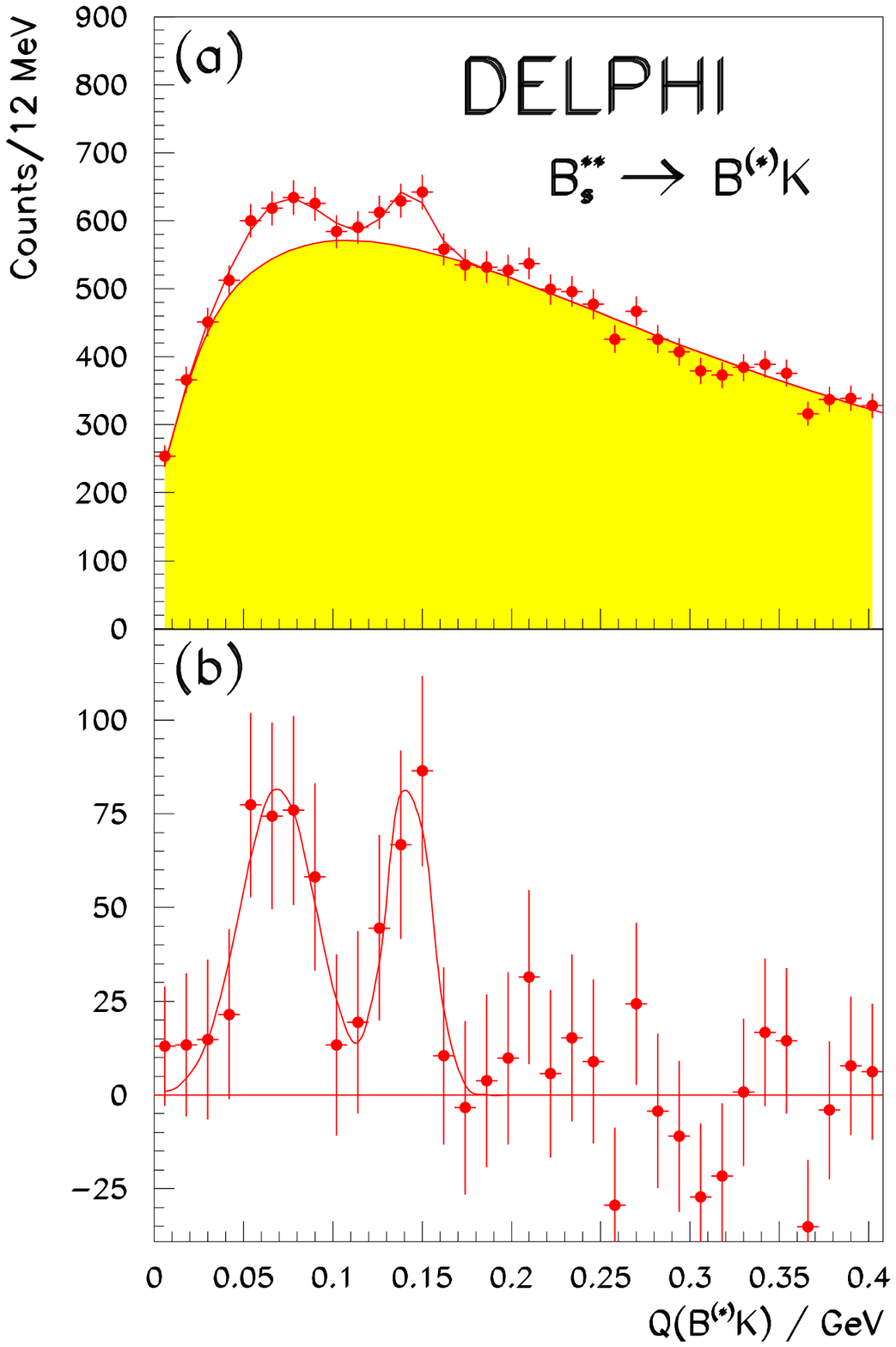}
\fcaption{(a) The distribution of the $Q$ value of inclusively reconstructed
$B$ hadrons and kaons observed by DELPHI. The points are the data, and the
shaded area is the predicted distribution from Monte Carlo simulation that
does not contain $B^{**}$;
(b) the resulting distribution when the predicted background from
Monte Carlo is subtracted from the data; the fit is described in the text.}
\label{fig:delphi_bsdstar}
\end{figure}

DELPHI and OPAL have also reported evidence of $B^{**}_{s}$.
The results are summarized in Table~\ref{tab:bs_dstar}.
In this case,
the charged particle that is combined with the inclusive $B$ is identified
as a kaon using either $dE/dx$ measurements or RICH (DELPHI only). 
Furthermore, DELPHI uses jet-charge flavour tagging to improve signal
to background. OPAL observes one peak, and DELPHI observes two narrow
peaks (see Figure~\ref{fig:delphi_bsdstar}), which they interpret as coming
from the $j_q=\frac{3}{2}$ doublet, $B^2_s$ and $B^1_s$. Given the experimental
resolution, the width of the peak at higher mass (presumably due to $B^2_s$) is
$1.5\sigma$ narrower than expected. 

\begin{table}[h] \centering
\tcaption{Summary of results on $B^{**}_{s}$. All results have been converted
to $Q=M(B\pi)-M(B)-M(K)$. The fitted parameters for a Gaussian are reported.
OPAL reports results from a single Gaussian fit, and DELPHI fits their
observed enhancement to two Gaussians. The production ratio is defined as
$\frac{B(Z\rightarrow b\rightarrow B^{**}_{s})}{B(Z\rightarrow b)}$.
For all reported measurements, the first error is statistical and the
second error is systematic.}
\label{tab:bs_dstar}
\small
\begin{tabular}{||c|c|c||}\hline\hline
Experiment                        & DELPHI & OPAL \\
\hline
Data Sample                       & $2.3\times10^6 Z^0$ (91--94)  &
                                    $3.5\times10^6 Z^0$ (91--94)   \\
\hline
Signal                            & $577\pm49\pm70$ &
                                    $149\pm31$       \\
\hline
$Q$ (MeV/$c^2$)                   & $70\pm4\pm8$ &
                                    $80\pm15$     \\
                                  & $142\pm4\pm8$ &
                                                   \\
\hline
$\sigma$ (MeV/$c^2$)              & $21\pm4\pm4$ &
                                    $36\pm5$      \\
                                  & $13\pm4\pm4$ &
                                                  \\
\hline
Prod.~ratio (\%)                  & $2.1\pm0.5\pm0.7$ &
                                    $2.1\pm0.4\pm0.5$  \\
                                  & $1.6\pm0.5\pm0.7$ &
                                                       \\
\hline
Reference                         & \cite{ref:delphi_eps0563} &
                                    \cite{ref:opal_bdstar}     \\ 
\hline\hline
\end{tabular}
\end{table}

\subsection{Results on Excited $b$~Baryons}

\begin{figure}[h] \centering
\mbox{}
\epsfysize 10cm
\epsffile{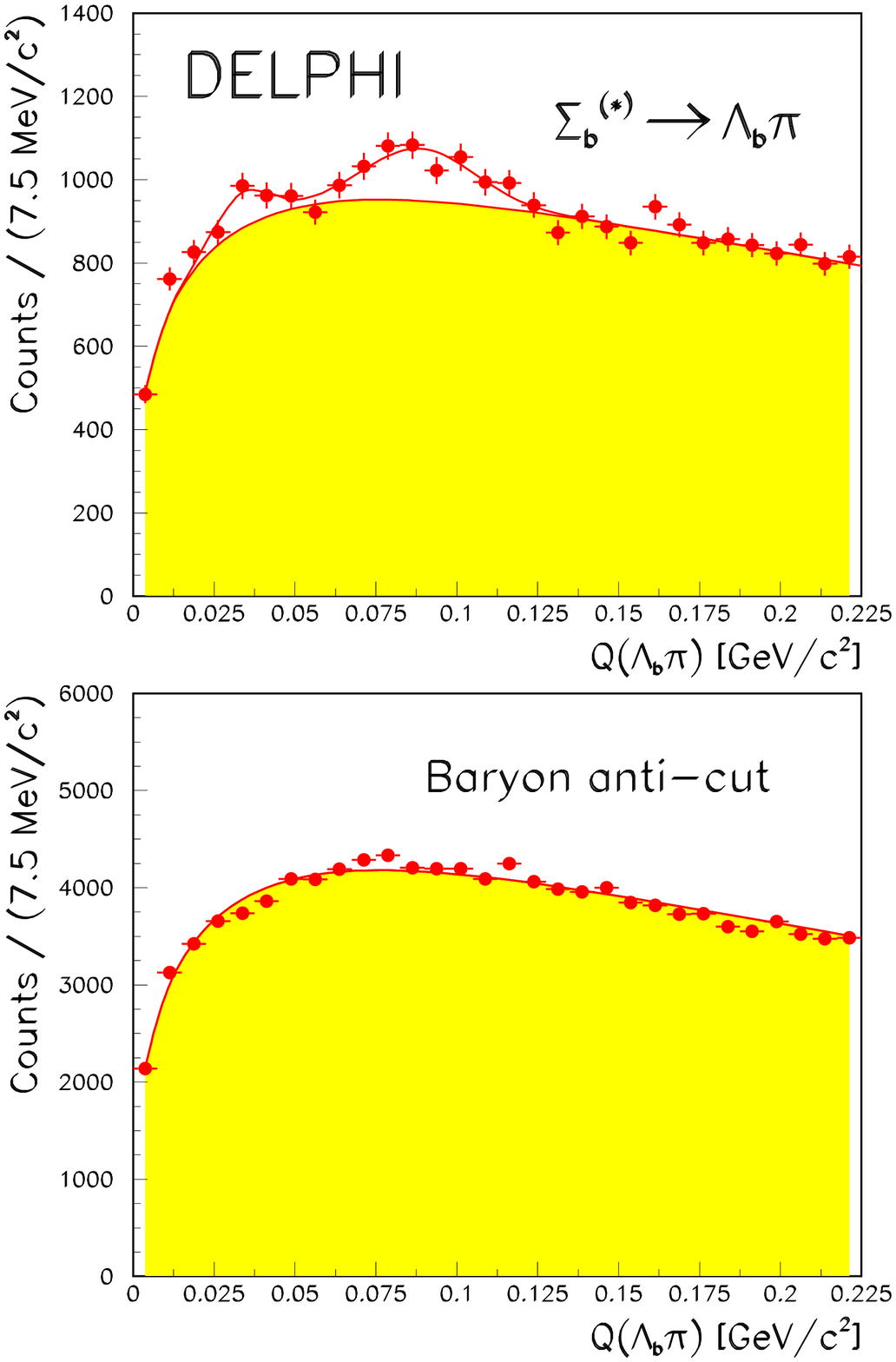}
\fcaption{(a) The distribution of $Q$ for inclusively reconstructed $B$~hadrons
and pions in a baryon enriched data sample from DELPHI. The points are the
data, and the shaded region is the expected background. The curve is the result
of a fit to the expected background shape and two Gaussians with widths
fixed to the predicted experimental resolution. (b) The same distribution
except the data are depleted in baryons; these data are used to determine
the background in (a).}
\label{fig:delphi_sigmab}
\end{figure}

DELPHI has reported\cite{ref:delphi_eps0565}
evidence for $\Sigma_b$ and $\Sigma^*_b$ decaying into
$\Lambda_b\pi$. They enhance the fraction of $\Lambda_b$ in their sample
of inclusively reconstructed $B$~hadrons by requiring that at least one
of the two most energetic particles in the hemisphere of the inclusive
$B$ be (1) an identified proton (using $dE/dx$ and RICH), (2) a $\Lambda$,
or (3) a neutral cluster in the hadronic calorimeter that exceeds 10 GeV in
energy. In this baryon-enriched sample, they find two enhancements in the
$Q=M(\Lambda_b\pi)-M(\Lambda_b)-M(\pi)$ distribution
(see Figure~\ref{fig:delphi_sigmab}). The characteristics
of these enhancements are summarized in Table~\ref{tab:delphi_sigmab}.
They reverse their baryon-enrichment requirements and find no evidence
of either signal (see Figure~\ref{fig:delphi_sigmab}); the data distribution
determined with these reverse criteria are used to determine the background.
They examine the helicity angle of the pion in the $\Sigma^{(*)}_b$
rest-frame and find an indication of suppression of $\pm\frac{3}{2}$~helicity
states. This suppression combined with significant production rates
of excited $b$~baryons could lead to a substantial
reduction of $\Lambda^0_b$ polarization\cite{ref:falk_peskin}.

\begin{table}[h] \centering
\tcaption{Summary of results on $\Sigma_b$ and $\Sigma^*_b$
reported\cite{ref:delphi_eps0565} by DELPHI.
The data sample is $2.3\times10^6$ hadronic $Z^0$ decays collected during
1991--1994. The parameters from fitting the two observed enhancements
with Gaussians are reported. The widths of the Gaussians were fixed to
the expected experimental resolution of 10 and 16 MeV/$c^2$.
The production ratio is defined as
$\frac{B(Z\rightarrow b\rightarrow\Sigma^{(*)}_b)}{B(Z\rightarrow b)}$.
The mass difference with the $\Lambda_b$ is obtained by adding the mass
of the charged pion to the measured $Q$ value.
For all reported measurements, the first error is statistical and the
second error is systematic.}
\label{tab:delphi_sigmab}
\small
\begin{tabular}{||c|c|c||}\hline\hline
Excited $B$ Baryon & $\Sigma_b$ & $\Sigma^*_b$ \\
\hline
Signal             & \multicolumn{2}{|c||}{$937\pm108\pm270$} \\
\hline
$Q$ (MeV/$c^2$)    & $33\pm3\pm8$ & $89\pm3\pm8$ \\
\hline
$\frac{\sigma(\Sigma_b)}{\sigma(\Sigma_b)+\sigma(\Sigma^*_b)}$ (\%)
                   & \multicolumn{2}{|c||}{$24\pm6\pm10$} \\
\hline
Prod.~ratio (\%)   & \multicolumn{2}{|c||}{$4.8\pm0.6\pm1.5$} \\
\hline
$M(\Sigma_b)-M(\Lambda_b)$ (MeV/$c^2$)
                   & $173\pm3\pm8$ & $229\pm3\pm8$ \\
\hline\hline
\end{tabular}
\end{table}

\subsection{$B^{**}$ from Exclusively Reconstructed $B$~Mesons}

ALEPH has reported\cite{ref:aleph_eps0403}
the observation of a $B\pi$ resonance using exclusively
reconstructed $B$~hadrons. To reconstruct the $B$ hadrons, they use a variety
of decay modes; most of the statistics come from the final states with
a $D$ or $D^*$ meson combined with  $\pi^{\pm}$, a $\rho^{\pm}$, or
an $a^{\pm}_1$. These same final states were used in the $\bar{B}^0$ and
$B^-$ lifetime measurements discussed previously, but the selection requirements
are relaxed to increase statistics.
The sample consists of 198 $\bar{B}^0$ candidates,
186 $B^-$ candidates, and an additional 90 $B^-$ candidates in which a
$\gamma$ or $\pi^0$ from $D^{*0}$ decay has not been detected. The purity
of the sample is estimated to be $(82\pm5)$\%.

Next they select candidate pions to combine with the $B$~hadron. The
selected track is required to be consistent with originating from the
interaction point and must have a measured $dE/dx$ consistent with a pion.
The charge of the pion from $B^{**}$ decay is correlated with $b$~quark
flavour: the combinations $B^-\pi^+$ and $\bar{B}^0\pi^-$ (and charge conjugate)
are right-sign combinations, and the combinations $B^-\pi^-$ and
$\bar{B}^0\pi^+$ are wrong-sign combinations. Resonant structure should
appear only in right-sign combinations; the wrong-sign combinations can
be used as one measure of the background.

Background pions come mainly from fragmentation.
To reduce this background, they choose
the pion candidate that has the maximum component of momentum 
projected on the $B$ candidate momentum.
Since the $b$~quark has a hard fragmentation function, pions from $B^{**}$
decay satisfy this requirement more often than pions produced in the
fragmentation of the $b$~quark.

\begin{figure}[h] \centering
\mbox{}
\epsfysize 7cm
\epsffile{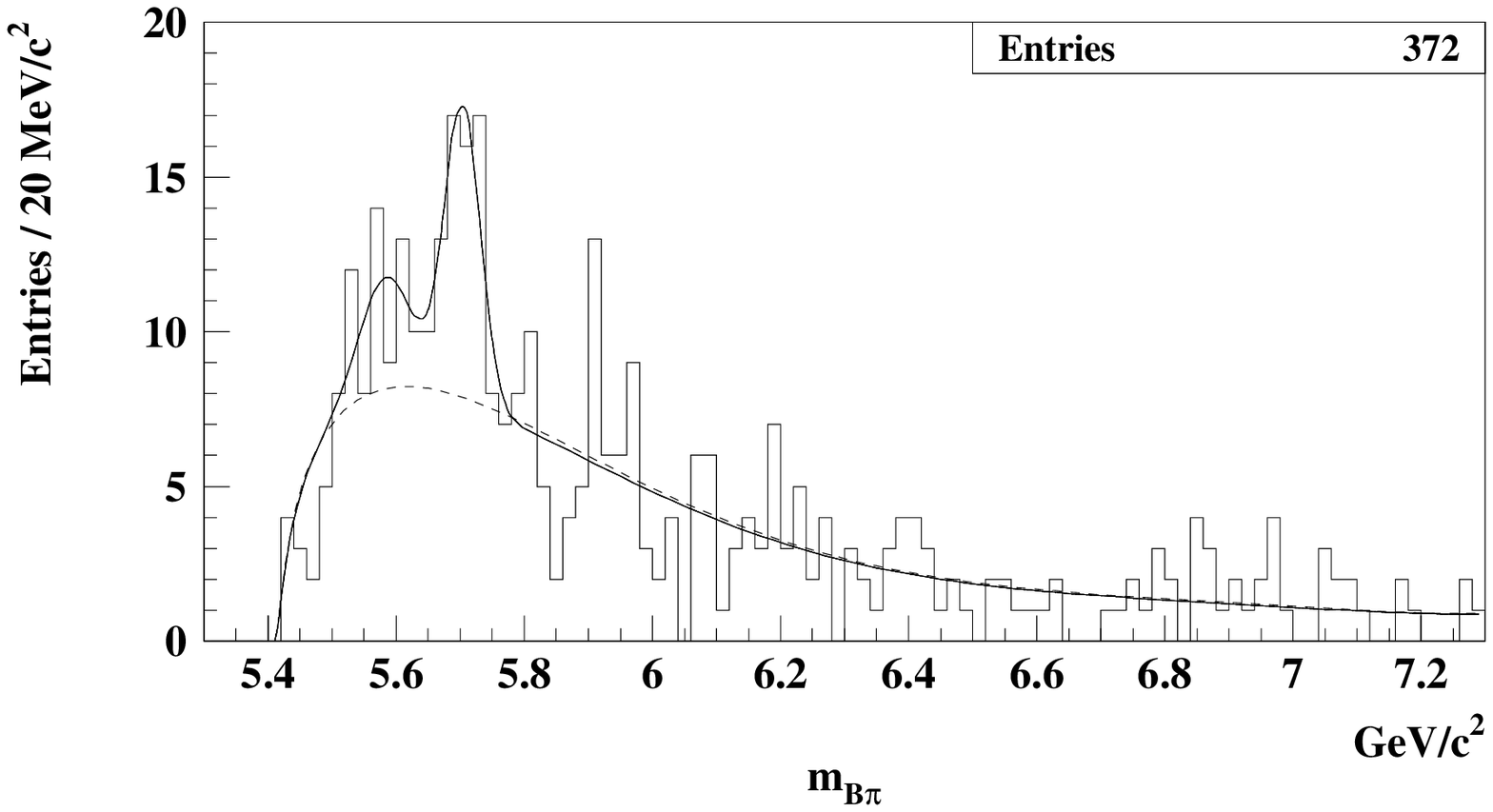}
\fcaption{The observed resonant structure of right-sign $B\pi$ combinations,
using exclusively reconstructed $B$~mesons in the ALEPH detector. The
data are the histogram; the dashed curve is the expected background; and
the solid curve is the combination of the background and two Gaussians
(see text for fitted parameters) used to describe the observed signal.}
\label{fig:aleph_bdstar}
\end{figure}

The observed resonant structure is shown in Figure~\ref{fig:aleph_bdstar}.
An unbinned maximum likelihood fit to two Gaussians yields the following
parameters:
$m_{narrow}=5703\pm14$ MeV/$c^2$, $\sigma_{narrow}=28^{+18}_{-14}$ MeV/$c^2$,
$m_{wide}=5585^{+79}_{-34}$ MeV/$c^2$, $\sigma_{wide}=42^{+43}_{-17}$ MeV/$c^2$,
and the total signal is $54^{+15}_{-14}$. The value of $m_{narrow}$ and
the observed production ratio $\frac{B(Z\rightarrow b\rightarrow B^{**}_{u,d})}
{B(Z\rightarrow b\rightarrow B_{u,d})}=(30\pm8)$\% are consistent with
the values measured using inclusive $B$ reconstruction.
The mass resolution using exclusive $B$ decays is 2 to 5 MeV/$c^2$, almost
an order of magnitude better that the resolution achieved with inclusive
$B$ reconstruction. 

\subsection{Results on Flavour Tagging}

ALEPH has used their exclusive $B$ sample, and OPAL has used their inclusive
$B^-$ sample to measure flavour tagging using the $B^{**}$ resonance and
associated fragmentation production. The effectiveness of a flavour tag
may be quantified by the product
$${\cal E}=\varepsilon\cdot(1-2w)^2,$$
where $\varepsilon$ is the efficiency of the flavour tag, and $w$ is the 
probability that the flavour assignment is incorrect
(the mistag probability); ${\cal E}$ is bounded between 0 and 1.
The error on an asymmetry $A$ measured with $N$ decays using a flavour
tag of quality ${\cal E}$ is
$$\delta A=\sqrt{\frac{1-A^2}{{\cal E}\cdot N}};$$
in other words, the equivalent statistics are reduced from $N$ to
${\cal E}\cdot N$.
Combining their samples of neutral and charged $B$~mesons,
ALEPH measures\cite{ref:aleph_eps0403} ${\cal E}=15\pm4$\%.
Using their sample of inclusively reconstructed charged $B$~mesons and
requiring the helicity angle of the pion satisfy $\cos\alpha_{\pi}>-0.7$,
OPAL measures\cite{ref:opal_bdstar} ${\cal E}=6.6\pm1.0$\% when they restrict
the mass of the $B\pi$ system to lie between $5.60<m_{B\pi}<5.85$ GeV/$c^2$,
and ${\cal E}=11\pm2$\% when no restriction to the resonant region is applied.
These tagging efficiencies compare favourably with other tags, for example,
the typical value of ${\cal E}$ using the lepton charge from semileptonic decays
of the other $B$ produced in the event is $\sim 2$\% (per lepton type).

\section{Summary}

A summary of the measured $B$~hadron lifetimes and lifetime-ratios is
given in Table~\ref{tab:lifesum}. The measured values of the ratios
$\tau(B^-)/\tau(\bar{B}^0)$ and $\tau(\bar{B}^0_s)/\tau(\bar{B}^0)$
agree with theoretical expectations. The measured value of
$\tau(\Lambda_b)/\tau(\bar{B}^0)$ is significantly below theoretical bounds.
The possible sources for this discrepancy are listed in
reference~[\cite{ref:matthias}].
Searches for the $B_c$~meson are negative thus far.

\begin{table}[h] \centering
\tcaption{A summary of the average of measured $B$~hadron lifetimes
and the ratio of these lifetimes with respect to $\tau(\bar{B}^0)$.}
\label{tab:lifesum}
\small
\begin{tabular}{||c|c|c||}\hline\hline
$B$ hadron & lifetime (psec) & ratio \\
\hline
$B^-$       & $1.62\pm0.05$ & $1.02\pm0.04$ \\
\hline
$\bar{B}^0$ & $1.56\pm0.05$ & --            \\
\hline
$\bar{B}^0_s$ & $1.55^{+0.11}_{-0.10}$ & $0.99\pm0.07$ \\
\hline
$\Lambda^0_b$ & $1.18\pm0.07$ & $0.75\pm0.05$ \\
\hline\hline
\end{tabular}
\end{table}

Excited $B$~hadrons have been reconstructed using inclusive reconstruction
of weakly decaying $B$~hadrons. $B^*$ production rates and polarization are
as expected from spin-counting.
The predicted p-wave mesons ($B^{**}$) provide a good
qualitative description of the observed resonant structure in the mass
distributions of $B\pi$ and $BK$ combinations. Flavour tagging of $B^0$ using
$B^{**}$ and associated non-resonant pion production appears promising.
Finally, the observed significant rate of $\Sigma^{(*)}_b$ may be part
of the reason that $\Lambda_b$ baryons are not fully polarized on the
$Z^0$~resonance.

The future of these measurements may be summarized as follows:
\begin{itemize}
\item{
{\em LEP} -- 1995 is the last year on the $Z^0$~resonance before the energy
upgrade. Each experiment will increase its data sample of hadronic $Z^0$
decays by at most $10^6$, yielding total hadronic samples of $4-4.5\times10^6$
per experiment. Except for specific analyses ({\it e.g.}~those exploiting
the RICH detector at DELPHI), adding the 1995 data sample to the results
submitted to this conference with not change the statistical errors
dramatically. Refinements of the analyses and the addition of further
decay channels will lead to further reduction of statistical and systematic
errors.}
\item{
{\em SLD} -- from 1996 to 1998, the SLC is expected to deliver approximately
$5\times10^5$ hadronic $Z^0$ decays with 80\% electron polarization.
This will triple the current SLD data sample. This increase in statistics,
along with a new CCD vertex detector with increased solid angle coverage
should lead to significant improvements in the measurements from SLD.}
\item{
{\em CDF} -- the results submitted to this conference came from either Run 1a
(20 pb$^{-1}$ taken during 1992--1993) or from Run 1a and part of
Run 1b (70 pb$^{-1}$).
Run 1b started in January 1994 and is expected to end in Spring of 1996.
The anticipated size of the recorded data sample of Run 1a and 1b combined
is 140 pb$^{-1}$. The results submitted to this conference will improve
after the full statistics are analysed (for example, the statistical error
on $\tau(B^-)/\tau(\bar{B}^0)$ should decrease from 0.09 to 0.07), and
many new results are anticipated. This fall, CDF has obtained three new
preliminary measurements. 
In addition to the measurement of the $\Lambda_b$ mass 
$m(\Lambda_b)=5623\pm5(stat.)\pm4(syst.)$~MeV/$c^2$, mentioned previously,
the $\Lambda_b$ lifetime has been measured\cite{ref:cdf_lbtau} to be
$\tau(\Lambda_b)=1.33\pm0.16(stat.)\pm0.07(syst.)$~psec using a signal
of $197\pm25$ $\ell\Lambda_c$ pairs. The increased statistics have
been exploited to improve the measurement of the $\bar{B}^0_s$ lifetime
using the exclusive decay $\bar{B}^0_s\rightarrow J/\psi\phi$.
From a signal of $58\pm8$ decays, they measure\cite{ref:cdf_bstau_update}
$\tau(\bar{B}^0_s)=1.34^{+0.23+0.05}_{-0.19-0.05}$~psec.
During Run II, which is anticipated to start in early 1999, after the main
injector in commissioned, the CDF and D0 collaborations should make definitive
measurements of the $B$~hadron lifetimes and lifetime ratios using
fully reconstructed decays.}
\end{itemize}

\section{Acknowledgements}

I would like to thank the organizers for a well-organized and enjoyable
symposium, and I thank Han Geng for the assistance he offered me.
I thank my fellow speakers Matthias~Neubert and
Tomasz~Skwarnicki for information provided before the symposium.
The following people have been very helpful in providing information
about the material presented in this summary:
Richard Batley (OPAL),
Bob Clare (L3),
Fritz DeJongh (CDF),
Michael Feindt (DELPHI),
Marta Felcini (L3),
Roger Forty (ALEPH),
Chris Hawkes (OPAL),
Richard Hawkings (OPAL),
Nikolaos Konstantinidis (ALEPH),
Alan Litke (ALEPH),
Luigi Rolandi (ALEPH),
Gary Taylor (ALEPH),
Daniel Treille (DELPHI),
Vivek Sharma (ALEPH),
Paris Sphicas (CDF),
Tracy Usher (SLD),
Rick Van~Kooten (OPAL),
Wilbur Venus (DELPHI),
David Ward (OPAL).
Special thanks to Robert Kowalewski (OPAL),
and to Hans-G\"{u}nter Moser (ALEPH), Lucia Di~Ciaccio (DELPHI)
and the LEP B lifetime working group.
These proceedings were carefully proofread by Manfred Paulini.
Special thanks to my wife Monica Kroll for her help in preparations
for the symposium.
This work is supported by the United States Department of Energy
under grant \#~DE-AC02-76CHO3000.

\section{References}
%It is preferred that references in the bibliography be cited in
%the text using a superscript number without parentheses or
%brackets; for example, if we cite the paper by Cohen and
%Anderson the reference number appears like this.~\cite{Coh/An}
%All references should include initials and last name of the
%author(s), title of publication (in italics), volume (in bold),
%year of publication of paper in the journal/book, and page
%numbers, e.g.,
%
In this list, ``contribution to this Symposium'' means a contribution
submitted to the XVII International Symposium on Lepton--Photon Interactions,
Beijing, 10--15 August 1995. If an EPS number follows, this number refers to
the contribution submitted to The International Europhysics
Conference on High Energy Physics, Brussels, 27 July 1995 to 2 August 1995.
% macros for references
\def\etal{{\it\space et al.},}
\def\issue(#1,#2,#3){{\bf{#1}},\space(#3)\space#2}
\def\PL(#1,#2,#3){ Phys.\ Lett.\ \issue(#1,#2,#3)}
\def\PLB(#1,#2,#3){ Phys.\ Lett.\ B\ \issue(#1,#2,#3)}
\def\NP(#1,#2,#3){ Nucl.\ Phys.\ \issue(#1,#2,#3)}
\def\NPB(#1,#2,#3){ Nucl.\ Phys.\ \ B\ \issue(#1,#2,#3)}
\def\NC(#1,#2,#3){ Nuovo Cimento\ \issue(#1,#2,#3)}
\def\NIM(#1,#2,#3){ Nucl.\ Instrum.\ \ Methods\ \issue(#1,#2,#3)}
\def\EL(#1,#2,#3){ Euro.\ Phys.\ Lett.\ \issue(#1,#2,#3)}
\def\ZP(#1,#2,#3){ Z.\ Phys.\ C\ \issue(#1,#2,#3)}
\def\PRL(#1,#2,#3){ Phys.\ Rev.\ Lett.\ \issue(#1,#2,#3)}
\def\PRD(#1,#2,#3){ Phys.\ Rev.\ D\ \issue(#1,#2,#3)}
\def\Arnison{The UA1 Collaboration, G.~Arnison\etal}
\def\Albajar{The UA1 Collaboration, C.~Albajar\etal}
\def\Abe{The CDF Collaboration, F.~Abe\etal}
\def\cdf{The CDF Collaboration}
\def\Buskulic{The ALEPH Collaboration, D.~Buskulic\etal}
\def\aleph{The ALEPH Collaboration}
\def\Abreu{The DELPHI Collaboration, P.~Abreu\etal}
\def\Adam{The DELPHI Collaboration, W.~Adam\etal}
\def\delphi{The DELPHI Collaboration}
\def\Adriani{The L3 Collaboration, O.~Adriani\etal}
\def\Acciarri{The L3 Collaboration, M.~Acciarri\etal}
\def\Akrawy{The OPAL Collaboration, M.~Z.~Akrawy\etal}
\def\Alex{The OPAL Collaboration, G.~Alexander\etal}
\def\Acton{The OPAL Collaboration, P.~D.~Acton\etal}
\def\Akers{The OPAL Collaboration, R.~Akers\etal}
\def\opal{The OPAL Collaboration}
\def\Kabe{The SLD Collaboration, K.~Abe\etal}
\def\lp{talk presented at the XVII International Symposium on Lepton--Photon
Interactions, Beijing, 10--15 August 1995, to appear in these proceedings}
\def\lponly{contribution to this Symposium}
\def\lpeps{contribution to this Symposium}
\def\subplb{submitted to Physics Letters B}
\def\subprd{submitted to Physical Review D}
\def\subzp{submitted to Zeitschrift f\"{u}r Physik C}


\begin{thebibliography}{9}
\bibitem{ref:hgm_private}
H. Moser, private communication; the only difference between the number quoted
here and the number cited in [\cite{ref:hgm}] is the updated number from
ALEPH\cite{ref:aleph_bstau} is used here.
\end{thebibliography}
\end{document}